\documentclass[%
reprint,
superscriptaddress,
nofootinbib,
 amsmath,amssymb,
 aps,
 pra,
longbibliography
]{revtex4-2}

\usepackage[english]{babel}
\usepackage{appendix}
\usepackage{graphicx}
\usepackage{dcolumn}
\usepackage{bm}
\usepackage{xcolor}
\usepackage{braket}
\usepackage{hyperref}
\usepackage[capitalise]{cleveref}
\newcommand{\sket}[1]{\ket{#1}\rangle}
\newcommand{\sbra}[1]{\langle\bra{#1}}

\newcommand\scalemath[2]{\scalebox{#1}{\mbox{\ensuremath{\displaystyle #2}}}}
\newcommand{\tr}{\text{Tr}}

\begin{document}

\preprint{APS/123-QED}

\title{Entanglement, information and non-equilibrium phase transitions \\
in long-range open quantum Ising chains}

\author{Daniel A. Paz}
\affiliation{%
 Department of Physics and Astronomy, Michigan State University, 426 Auditorium Rd. East Lansing, MI 48823, USA
}
\affiliation{Kavli Institute for Theoretical Physics, University of California, Santa Barbara, CA 93106-4030, USA}

\author{Benjamin E. Maves}
\affiliation{%
 Department of Physics and Astronomy, Michigan State University, 426 Auditorium Rd. East Lansing, MI 48823, USA
}
\affiliation{Department of Physics, Syracuse University, Syracuse, NY 13244, USA}

\author{Naushad A. Kamar}
\affiliation{%
 Department of Physics and Astronomy, Michigan State University, 426 Auditorium Rd. East Lansing, MI 48823, USA
}
\affiliation{Kavli Institute for Theoretical Physics, University of California, Santa Barbara, CA 93106-4030, USA}

\author{Arghavan Safavi-Naini}
\affiliation{Kavli Institute for Theoretical Physics, University of California, Santa Barbara, CA 93106-4030, USA}
\affiliation{Institute for Theoretical Physics, Institute of Physics, University of
Amsterdam, Science Park 904, 1098 XH Amsterdam, the Netherlands}

\author{Mohammad Maghrebi}
\email{maghrebi@msu.edu}
\affiliation{%
 Department of Physics and Astronomy, Michigan State University, 426 Auditorium Rd. East Lansing, MI 48823, USA
}
\affiliation{Kavli Institute for Theoretical Physics, University of California, Santa Barbara, CA 93106-4030, USA}
\affiliation{Pritzker School of Molecular Engineering, University of Chicago, Chicago, IL 60637}

\date{\today}

\begin{abstract}
Non-equilibrium phase transitions of open quantum systems generically exhibit diverging classical but not quantum correlations. Still entanglement---characterizing the latter correlations---can be sensitive to the phase transition. Furthermore, mutual information, bounding the total correlations, should exhibit critical scaling at the transition. In this work, we study these quantities in the steady state of open quantum Ising chains with power-law interactions (with the exponent $0\le \alpha \le 3$) where spins are subject to spontaneous emission. The bulk of this paper is dedicated to a detailed analytical as well as numerical analysis of the infinite-range model ($\alpha=0$), a model that is closely related to the paradigmatic open Dicke model. Our main findings are that the entanglement, while being finite, peaks, exhibits a kink and takes a universal value at the transition, while the mutual information exhibits critical scaling not only at the transition but well into the ordered phase, underscoring a hidden criticality that is not captured by (two-point) correlations. We consider three distinct entanglement measures: logarithmic negativity; quantum Fisher information; and, spin squeezing.
Specifically, we show that the collective spin operator that maximizes the quantum Fisher information can be identified with the \textit{gapless} mode of the phase transition,
while the squeezed direction is that of the \textit{gapped} mode.
Finally, we investigate power-law interacting models using matrix product states 
where we find comparable bounds on squeezing even when no phase transition is expected (for larger $\alpha$), thus the connection to the phase transition does not appear to hold for shorter-range interactions.

\end{abstract}

\maketitle

\section{Introduction}
Quantum entanglement is a defining signature of quantum mechanics, characterized by a nontrivial superposition of multi-particle states, and has been a subject of intense research in the past decades \cite{horodecki_quantum_2009}. Additionally, it has been shown to be useful as a resource for quantum computation \cite{ding_review_2007, huang_superconducting_2020, georgescu_quantum_2014} and metrology \cite{giovannetti_quantum_2006, giovannetti_advances_2011}. However, entanglement is delicate, and is generically spoiled in the presence of noise or coupling to the environment. A standard solution is to further isolate the system and to cool it down to lower temperatures to minimize dissipation and thermal fluctuations. An attractive alternative that has become more popular recently is to utilize engineered dissipation \cite{harrington2022engineered} to create entangled states \cite{krauter_entanglement_2011, zippilli_entanglement_2013, lin_dissipative_2013, Wang_2013, liu_comparing_2016, boneberg_quantum_2021,Doucet_2023}. In this work, we  rather investigate 
open quantum systems in the presence of generic dissipation, a.k.a. driven-dissipative systems, without any engineering and study entanglement as well as information. 
Specifically, our goal is to identify if and how such quantities are sensitive to disorder-to-order phase transitions of many-body driven-dissipative systems. 

These systems however pose a challenge as the equilibrium toolbox is not immediately available. In addition, entanglement and information being nonlinear in the density matrix are further nontrivial to compute analytically. Even numerically, open quantum system require a larger (vectorized) Hilbert space, and an exact numerical simulation is thus challenging. This motivates the first part (and the bulk) of our paper where we study the infinite-range driven-dissipative Ising model (iDDIM) \cite{paz_driven-dissipative_2021, paz_time-reversal_2021}, a  model that is closely related to the paradigmatic open Dicke model \cite{baden_realization_2014, baumann_dicke_2010, baumann_exploring_2011, nagy_critical_2011, torre_keldysh_2013, kirton_suppressing_2017, damanet_atom-only_2019, boneberg_quantum_2021} and exhibits an Ising phase transition from the normal (or, disordered) to an ordered phase.  
Owing to its infinite range, this model is accessible both analytically using covariance matrix techniques and numerically using quantum trajectories combined with the permutation symmetry. 
We calculate three distinct entanglement measures, namely the logarithmic negativity $E_{\mathcal{N}}$, the quantum Fisher information $F$ with applications in quantum metrology, and the spin squeezing parameter $\xi$, all proper entanglement measures even for mixed states. 
Furthermore, we calculate the von Neumann entropy, purity and mutual information; the latter captures both classical and quantum correlations in a mixed state and provides a useful contrast against true entanglement measures. Finally, we numerically study the long-range power-law variant of iDDIM with an exponent $\alpha \le 3$  using matrix product states and focus specifically on the squeezing parameter. \par

Before we continue, we provide a brief summary of our
main results. 
We find that the entanglement, while being finite everywhere in the phase diagram, peaks, features a kink, and takes a universal value at the phase boundary. 
Specifically, the logarithmic negativity takes the maximal value of $E_\mathcal{N} = 1/2$ in units of $\log 2$ at the transition. We also define the quantum Fisher information density $f_{\mathbf{n}} = F_\mathbf{n} /N$ (with $N$ the number of spins) corresponding to the component of the collective spin operator along the axis $\mathbf{n}$.
We show that this quantity is bounded from above, $f_{\mathbf{n}} \leq 2$, and the bound is only saturated at the transition and for $\mathbf{n}$ parallel to the gapless mode of the phase transition
\cite{paz_time-reversal_2021}. 
We note that a value greater than unity indicates that the state is at least 2-particle entangled. 
Similarly, the steady state is most squeezed at the phase boundary with the squeezing parameter $\xi = 1/2$ where the squeezed component is along the gapped mode, orthogonal to the gapless mode, of the phase transition.
This behavior is consistent with the 3 dB limit of parametric amplifiers at their threshold, underscoring the similarities between this threshold behavior and the driven-dissipative phase transition.
The most surprising finding of our work is  that the mutual information exhibits logarithmic scaling with system size not only at the phase boundary, but everywhere in the ordered phase albeit with a different (typically larger) coefficient than the universal value at the transition. This underscores a hidden criticality that cannot be accounted for by (two-point) correlations.  
Finally, for long-range power-law Ising models with the exponent $\alpha\le 3$, we numerically find that the squeezing is still bounded by $\xi = 1/2$, but the connection to phase transition is absent for large $\alpha$ where there is no phase transition.

This paper is structured as follows. In Sec.~\ref{sec model}, we introduce the long-range driven-dissipative Ising model and discuss the mean-field phase diagram. We  then focus on the infinite-range model (iDDIM) before coming back to the long-range model towards the end. In Sec.~\ref{sec cov mat}, we briefly introduce covariance matrices, and report explicit expressions for the covariance matrix in the normal phase.
Section~\ref{Entropy and Purity} discusses the von Neumann entropy, purity and the mutual information throughout the phase diagram.
Section~\ref{sec:ent_measures} is dedicated to quantifying different entanglement measures: we discuss the logarithmic negativity in Sec.~\ref{sec log neg}, and provide an analytical expression in the normal phase; in Sec.~\ref{sec QFI}, we calculate the quantum Fisher information density along the three main directions of the total spin, and identify the optimal direction analytically within the normal phase; we further calculate the spin squeezing parameter in Sec.~\ref{sec squeezing} analytically in the entire phase diagram.
On top of analytical results, extensive numerical results are provided using quantum trajectories and by taking advantage of the permutation symmetry of the iDDIM.  
We introduce the long-range Ising model in Sec.~\ref{sec_long_range_Ising} and compute the squeezing parameter numerically using matrix product states techniques. Finally, we summarize our results and discuss future directions in Sec.~\ref{sec:conclusion}. 
 
\section{Model}\label{sec model}
We consider $N$ coherently driven 2-level atoms in a transverse field $\Delta$ and subject to power-law Ising interactions
\begin{equation}\label{eq:full_Hamiltonian}
    H = -\frac{1}{\cal N}\sum_{i\ne j}\frac{\sigma_{i}^{x}\sigma_{j}^x }{|i-j|^\alpha}+ \Delta\sum_{i}\sigma_{i}^z ,
\end{equation}
with $\alpha$ the exponent of long-range interactions. Here, we have defined the normalization (Kac) factor,  
\(
    \mathcal{N} = \frac{1}{N-1}\sum_{i\ne j}|i-j|^{-\alpha},
\)
for convenience;
it renders the Hamiltonian extensive for $\alpha\le 1$ and furthermore simplifies the form of the mean-field equation (see below) for all values of $\alpha$.
Specifically, for $\alpha=0$, the Hamiltonian becomes
 an all-to-all Ising model (up to an unimportant additive constant):
\begin{equation}\label{Hamiltonian} 
     H = -\frac{J}{N} S_x^2 + \Delta  S_z\,,
\end{equation}
where we have defined the collective spin operators $S_a = \sum_i \sigma^a_i$ with $a \in \{x,y,z\}$. The normalization factor of $1/N$ renders the Hamiltonian extensive.  

Beside the Hamiltonian dynamics, atoms are coupled to a zero-temperature bath resulting in the spontaneous emission at a rate $\Gamma$. Assuming the standard Born-Markov approximation, the dynamics is governed by the quantum master equation in the Lindblad form:
\begin{equation}\label{iDDIM}
    \frac{\partial  \rho}{\partial t} = \mathcal{L}[ \rho] = -i[ H,  \rho] + \Gamma \sum_i  \sigma^-_i  \rho  \sigma^+_i - \frac{1}{2}\{ \sigma^+_i  \sigma^-_i,  \rho\}\,.
\end{equation}
We emphasize that this model describes a driven-dissipative system, although there is no explicit time dependence in the rotating frame. For $\alpha=0$, the resulting model, also referred to as the iDDIM, is a direct descendant of the open Dicke model in the large-detuning limit, as shown by some of the authors as well as others \cite{morrison_dynamical_2008, damanet_atom-only_2019, paz_driven-dissipative_2021}; see also \cite{dimer_proposed_2007}. Specifically, the Ising interaction strength arises due to the drive at the microscopic level. Closely related models have been recently proposed in the context of trapped ions as well \cite{sierant_dissipative_2021,Haack_2023}; see also related model with long-range dissipation \cite{Seetharam_2022}.
The above models exhibit a $\mathbb{Z}_2$ symmetry under the transformation $\sigma^{x,y} \to -\sigma^{x,y}$. 
For sufficiently small $\alpha$, this symmetry is spontaneously broken in the non-equilibrium steady state of \cref{iDDIM} exhibiting a transition from the normal phase ($\langle {S}_{x/y} \rangle = 0$) to the ordered phase ($\langle {S}_{x/y} \rangle \neq 0$). Below, we provide a mean-field analysis which becomes exact for $\alpha =0$. 
\par

\subsection{Mean field}\label{sec MF}
\begin{figure}
     \centering
     \includegraphics[width=\linewidth]{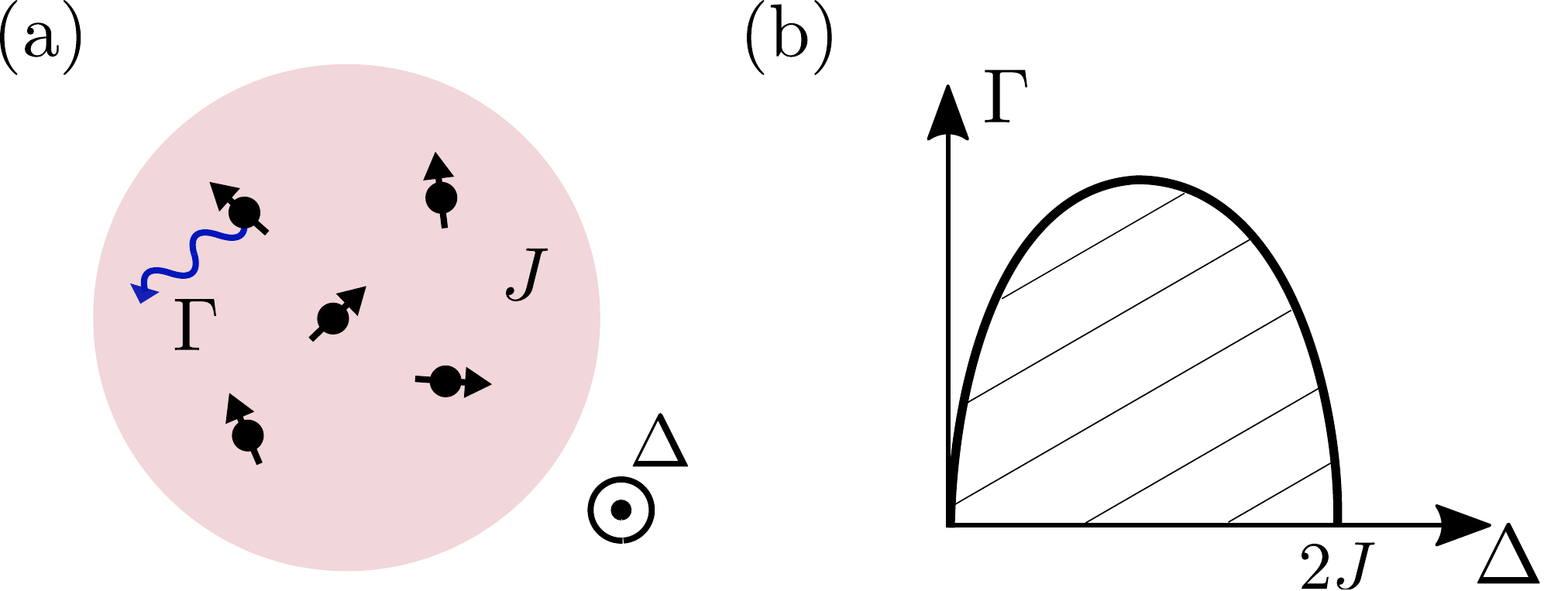}
     \caption{(a) Schematic depiction of the driven-dissipative Ising model with infinite-range interactions. (b) Phase diagram of the model. The shaded region designates the ordered phase.}
     \label{fig phase diagram}
 \end{figure}

We briefly discuss the mean-field solution of Eq.~\eqref{iDDIM}. We first assume that the density matrix factorizes as
\begin{equation}
    \rho = \bigotimes_i \rho_i = \rho_{\text{MF}}^{\otimes N}\,,
\end{equation}
where $\rho_{\text{MF}}$ is the mean-field density matrix, uniform across all sites. 
Inserting this ansatz in the equations of motion for $\sigma^{x/y/z}$, we find the mean-field equations of motion (in the $N \to \infty $ limit)
\begin{subequations}
\begin{align}
    \partial_t \langle \sigma^x \rangle &= -2\Delta \langle \sigma^y \rangle - \frac{\Gamma}{2}\langle \sigma^x\rangle\,, \\
    \partial_t \langle \sigma^y \rangle& = (4J \langle \sigma^z \rangle + 2\Delta )\langle \sigma^x \rangle - \frac{\Gamma}{2} \langle \sigma^y \rangle\,, \\
    \partial_t \langle \sigma^z \rangle &= -4J \langle \sigma^y \rangle \langle \sigma^x \rangle - \Gamma(1 + \langle \sigma^z \rangle)\,,
\end{align}
\end{subequations}
where we have dropped the spatial index due to the uniform ansatz. Notice that with our convenient choice of the Kac factor, the dependence on $\alpha,N$ drops out of these equations. We seek the steady state solutions of the equations of motion. In the normal phase, there is only one stable solution, $\langle \sigma^x \rangle_{\text{ss}} = 0, \langle \sigma^y \rangle_{\text{ss}} = 0, \langle \sigma^z \rangle_{\text{ss}} = -1$ with the subscript indicating the steady state. In the ordered phase, there are now two stable solutions,
\begin{subequations}\label{mean-field eqs}
\begin{align}
    \langle \sigma^x \rangle_{\text{ss}} &= \pm \frac{\sqrt{32 J \Delta - 16\Delta^2 - \Gamma^2}}{4 \sqrt{2}J}\,, \\
   \langle \sigma^y \rangle_{\text{ss}} &= \mp \frac{\Gamma \sqrt{32 J \Delta - 16\Delta^2 - \Gamma^2}}{16 \sqrt{2}J \Delta}\,,  \\
    \langle \sigma^z \rangle_{\text{ss}} &= -\frac{\Gamma^2 + 16\Delta^2}{32J\Delta}\,.
\end{align}
\end{subequations}
The mean-field phase boundary is then determined as
\begin{equation}
    \Gamma^2 + 16\Delta^2 - 32J\Delta = 0\,,
\end{equation}
which can be understood as a phase transition from a normal (disordered) phase at large $\Gamma$ to a magnetically ordered phase for  $\Gamma \le \Gamma_c = 4\sqrt{\Delta(2J-\Delta)}$; see Fig.~\ref{fig phase diagram}(b). For a collective model with $\alpha=0$ and long-range models with $\alpha\le 1$ (i.e., with the exponent smaller than dimensionality), the mean-field phase diagram should be exact although (Gaussian) fluctuations on top of the mean-field solution plays an important role (see below). However, for sufficiently large $\alpha$, no phase transition is expected, as we shall discuss later, and the mean-field analysis just provides a first guide.  

\subsection{Gaussian Fluctuations}\label{sec gaussian}
On top of the mean-field solution in Eq.~\eqref{mean-field eqs}, we should also consider quantum and statistical fluctuations. 
A possible avenue is to cast the Liouvillian dynamics in terms of functional integral techniques using the approach in previous works \cite{paz_driven-dissipative_2021,paz_time-reversal_2021}.
Specifically for $\alpha=0$ (more generally, for any $\alpha<1$), one needs to include only fluctuations up to the quadratic order beyond the mean-field solution, at least away from the phase transition. 
These fluctuations can be then completely characterized by two-point correlation functions, that is, they are fully characterized by a Gaussian state. 
We shall leave the technical details to \cref{split system appendix,ordered phase appendix} and use the correlations reported there. A similar functional integral approach can be extended to nonzero $\alpha$ (including $\alpha$ larger than, but close to, 1). However, we shall not pursue the field-theory approach for $\alpha>0$ in this work. \par 

In the remaining of this paper, we first exclusively focus on the iDDIM (with $\alpha=0$) and only in \cref{sec_long_range_Ising} we report our results for $\alpha\le 3$.

\section{Covariance Matrix Method}\label{sec cov mat}
Since the fluctuations in the iDDIM are purely Gaussian at large system sizes \cite{paz_driven-dissipative_2021}, the state can be fully characterized by two-point correlation functions. More formally, a Gaussian state can be fully parameterized by a displacement vector $\mathbf{d}$ plus a covariance matrix $\sigma$,
\begin{equation}
    d_i = \tr(\rho r_i)\,, \qquad 
    \sigma_{ij} = \tr(\rho \{\Delta r_i, \Delta r_j\})\,,
\end{equation}
where $\mathbf{r} = (x_1, x_2,..., p_1, p_2, ...)^T$ is the vector comprising position and momentum operators, and $\Delta \mathbf{r} = \mathbf{r} - \mathbf{d}$; the curly brackets denote the anti-commutator. We calculate  entanglement measures using the covariance matrix and its \textit{symplectic} eigenvalues \cite{peschel_calculation_2003, serafini_symplectic_2003, adesso_extremal_2004, peschel_reduced_2009, eisler_entanglement_2014, safranek_optimal_2016}. These eigenvalues are obtained by diagonalizing the matrix $i \Omega \sigma$, where $\Omega$ is defined as 
\begin{equation}
    \Omega = \begin{pmatrix} 0 & I \\ -I & 0 \end{pmatrix}\,,
\end{equation}
and $I$ is the identity matrix; the matrix $\Omega$
encodes the canonical commutation relations of the position and momenta operators:
\begin{equation}
    [r_i, r_j] = i \Omega_{ij}\,.
\end{equation}
The eigenvalues of $ i \Omega \sigma$ always come in pairs of $\pm \nu_i$ and are bounded from below by 1. In the following subsections, we discuss how the covariance matrix formalism applies to the driven-dissipative Ising model in the normal phase while the ordered phase requires a different treatment. \par 

\subsection{Normal Phase}
A key observation in the normal phase is that the total spin is fully polarized along the negative $z$-direction with sub-extensive  fluctuations. Therefore, we can readily make the approximation $[S_x, S_y] = 2iS_z \approx  -2iN$, and then rescale the spin operators, $x = S_{x}/\sqrt{2N}, p = -S_y/\sqrt{2N}$, such that we retrieve the canonical commutation relations $[x, p] = i$. This identification gives a single pair of position and conjugate momentum operators, resulting in a single mode covariance matrix with $\mathbf{r} = (x, p)$.\par

The correlation functions for the steady state of Eq.~\eqref{iDDIM} in the normal phase have been computed in a previous work \cite{paz_time-reversal_2021}. 
We refer the reader to Appendix \ref{split system appendix} for the technical details and only quote the exact analytical expressions for the corresponding correlation functions: 
\begin{align}\label{covariance mat}
\begin{split}     
     \sigma_{11} &=  1+ \frac{16 J \Delta}{\Gamma^2 - \Gamma_c^2}\,, \\ 
     \sigma_{12}  &= \sigma_{21} =  \frac{4J\Gamma}{\Gamma^2 - \Gamma_c^2} \,, \\ 
     \sigma_{22}  &= 1+ \frac{ 16 J (2J - \Delta)}{\Gamma^2 - \Gamma^2_c}  \,.
\end{split}
\end{align}
All desired quantities, with the exception of the logarithmic negativity and mutual information, can be calculated using the covariance matrix representation of the density matrix. The logarithmic negativity and mutual information require that that we split the system in two and compute correlation functions between the two halves. The corresponding covariance matrix, denoted by $\sigma^{AB}$ for two subsystems $A$ and $B$, becomes a $4 \times 4$ matrix given by
\begin{equation}\label{split cov mat}
    \sigma^{AB} = \begin{pmatrix}
    X & K \\ K & P
    \end{pmatrix}\,,
\end{equation}
with the block matrices
\begin{align}
      X &= I  +  \frac{8 J \Delta}{\Gamma^2 - \Gamma_c^2} \begin{pmatrix} 1 & 1 \\ 1 & 1\end{pmatrix}\,, \nonumber \\ 
       P &= I  + \frac{8 J (2J -\Delta) }{\Gamma^2 - \Gamma_c^2} \begin{pmatrix} 1 & 1 \\ 1 & 1\end{pmatrix}\,, \label{split cov matrix elements}\\ 
       K &= \frac{2J \Gamma}{\Gamma^2 - \Gamma_c^2} \begin{pmatrix} 1 & 1 \\ 1 & 1\end{pmatrix} \nonumber\,.
\end{align}
For a derivation, see Appendix \ref{split system appendix}. We emphasize that these exact expressions are only valid in the normal phase.  \par

\subsection{Ordered Phase}\label{cov mat ordered phase}
The covariance matrix techniques discussed in the previous subsection apply only to bosonic systems with Gaussian fluctuations. 
In the normal phase, the iDDIM satisfies this condition as the total spin is fully polarized in the $z$-direction, and excitations of the spin can be seen as excitations of a bosonic mode akin to a Holstein-Primakoff transformation  \cite{das_infinite-range_2006, torre_keldysh_2013}. However, this picture no longer applies in the ordered phase. While we can always rotate our spin variables such that the spin is pointing along the $z$-direction, 
the spin is not fully polarized along the latter direction and fluctuations are generically non-negligible along all spin directions. 
While the covariance method is not immediately applicable in the ordered phase (however, see \cite{boneberg_quantum_2021}), we utilize quantum trajectory simulations to numerically calculate the desired entanglement measures in the ordered phase.

\section{Entropy, Purity and Information}\label{Entropy and Purity}
The von Neumann entropy 
\begin{equation}
    S_{\text{vN}}(\rho) = -\tr(\rho \log \rho)\,,
\end{equation}
defines the entropy of a given (mixed) state. While it is not a measure of entanglement in a mixed state, 
it can predict nontrivial behavior, especially at criticality, and is useful for comparison with other quantities. In this section, we analytically calculate the von Neumann entropy $S_{\text{vN}}$ both in the normal and ordered phases.

The von Neumann entropy in the normal phase can be calculated purely in terms of the symplectic eigenvalues of the covariance matrix. The mean-field contribution to the entropy is zero since the mean-field state, being fully polarized along the negative $z$ directions, is pure in the normal phase. It is known that $S_{\text{vN}}$ for a Gaussian state is given by \cite{adesso_extremal_2004} 
\begin{equation}\label{vN entropy}
    S_{\text{vN}}(\rho) = \frac{\nu +1 }{2}\log\left( \frac{\nu + 1}{2}\right) - \frac{\nu - 1 }{2}\log \left( \frac{\nu - 1}{2} \right)\,,
\end{equation}
where $\nu$ is the symplectic eigenvalue of $\sigma$. Equipped with the covariance matrix, given by Eq. \eqref{covariance mat}, we have
\begin{equation}\label{total system symplectic eigenvalue}
    \nu = \sqrt{1 + \frac{16J^2}{\Gamma^2 - \Gamma_c^2}}\,.
\end{equation}
This eigenvalue satisfies $\nu \geq 1$ in the normal phase $\Gamma > \Gamma_c$. Specifically, in the limit $J \to 0$ we have $\nu = 1$ and $S_{\rm vN} =0$, consistent with the state being pure and all spins pointing down \cite{adesso_extremal_2004}. 
More generally, the above equations give the von Neumann entropy everywhere in the normal phase. From the form of the symplectic eigenvalue, we can see that the von Neumann entropy diverges at $\Gamma = \Gamma_c$, signifying the onset of criticality. This is expected as fluctuations also diverge at the critical point \cite{paz_driven-dissipative_2021}. This phase transition was shown to be in the same university class as the finite-temperature transverse-field Ising model \cite{paz_driven-dissipative_2021} where the transition is dominated by thermal fluctuations. Thus, we expect $S_{\text{vN}}$ to diverge as it captures classical fluctuations as well. More precisely the von Neumann entropy diverges upon approaching the critical point as $S_{\text{vN}} \sim -\frac{1}{2}\log\gamma $ with $\gamma \equiv \Gamma - \Gamma_c$. Using the finite-size scaling analysis (see \cite{paz_driven-dissipative_2021, paz_driven-dissipative_2022}), we can substitute  $\gamma \sim 1/\sqrt{N}$ at the critical point, which in turn gives 
\begin{equation}
    S_{\text{vN}} \sim \frac{1}{4}\log N\,.
\end{equation}
The coefficient, $c=1/4$, of the logarithm is distinct from the zero-temperature equilibrium value of $1/6$ \cite{barthel_entanglement_2006}. This behavior is also distinct from expected the volume law behavior (with $S_{\rm vN} \propto N$) at finite temperature. The mutual information, however, behaves similarly $(\sim \frac{1}{4}\log N)$ as that of the Ising model at finite temperature \cite{wilms_finite-temperature_2012}, as we shall discuss shortly.

\begin{figure}
    \hspace*{-0.5cm} 
    \includegraphics[scale=0.5]{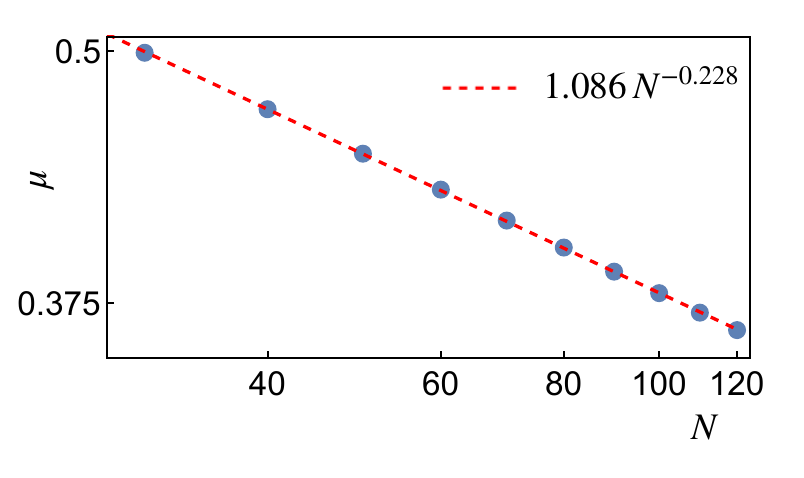}
     \caption{Algebraic scaling of the purity at the phase transition. The scaling is consistent with the theoretical prediction $\sim N^{-1/4}$.} 
     \label{fig:purity}
\end{figure}

To complement the von Neumann entropy, we can also calculate the purity of the state $\mu = \tr(\rho^2)$. In terms of the covariance matrix, the purity is given by
\begin{equation}
    \mu = \frac{1}{\sqrt{\text{det} (\sigma)}}\,.
\end{equation}
Conveniently, the symplectic eigenvalue of a single-mode covariance matrix is directly related to the determinant of the matrix, $\text{det}(\sigma) = \nu^2$, hence
\begin{equation}
    \mu = \frac{1}{\nu}\,.
\end{equation}
One can the see that, within the normal phase, the purity $\mu \propto \sqrt{\gamma}$
vanishes upon  approaching the critical point where the classical fluctuations diverge. Invoking the finite-size scaling again, we have 
\begin{equation}
    \mu \propto N^{-1/4},
\end{equation}
at the critical point. Indeed, we find a consistent behavior numerically; see \cref{fig:purity}. The numerical result is computed by vectorizing the  density matrix $\rho \to |\rho\rangle\!\rangle$, taking advantage of the permutation symmetry \cite{paz_driven-dissipative_2021}, and computing $\mu=\langle\!\langle|\rho|\rho\rangle\!\rangle$. 

As discussed in Sec. \ref{cov mat ordered phase}, we cannot perform the same procedure in the ordered phase. 
However, a simple analysis shows that the entropy is dominated by the mean-field contribution which is proportional to  $N$, hence a \textit{volume} scaling. This is simply due to the mean-field contribution as
\begin{equation}
    S_{\text{vN}}(\rho_{\text{MF}}^{\otimes N}) = N\, S_0(\rho_{\text{MF}})\,.
\end{equation}
While the entropy does transition from ``area law'' ($S_{\rm vN} \sim O(1)$) in the normal phase to ``volume law'' in the ordered phase, this behavior is distinct from the entanglement phase transitions at the level of individual quantum trajectories as opposed to the density matrix \cite{li_measurement-driven_2019, gullans_dynamical_2020, ippoliti_entanglement_2021, ippoliti_postselection-free_2021}. The volume law observed here simply reflects the mixed-ness of the state, while the area law in the normal phase should be attributed to the infinite-range interactions; the steady state of a generic (e.g., short-range) driven-dissipative system will be generically  mixed.

We also remark that the steady state becomes increasingly mixed in the ordered phase. The mean-field prediction of the purity is $\mu(\rho_{\text{MF}}) = \mu_0^N$ where $\mu_0=\frac{1}{2}(1 + s^2) < 1$, with $s = S/N = \sqrt{\langle S_x \rangle^2 + \langle S_y \rangle^2 + \langle S_z \rangle^2 }/N$. From the mean-field equation, Eq.~\eqref{mean-field eqs}, we then find 
\begin{equation}\label{purity}
    \mu_0 = 1- \frac{(\Gamma^2 - \Gamma_c^2)^2}{2048 J^2 \Delta^2}\,,
\end{equation}
for $\Gamma < \Gamma_c = 4\sqrt{\Delta(2J-\Delta)}$.
Since $\mu_0 < 1$ in the ordered phase, the mean-field solution predicts a mixed steady state. In fact, the total purity falls off exponentially with system size in the ordered phase, to be contrasted with the algebraic decay at the critical point. We also note that the  purity reaches its minimum when $\Gamma, \Delta \to 0$ within the ordered phase (with the order $\Gamma\to 0$ and then $\Delta\to0$) in which limit $\mu_0 = 1/2$ consistent with a fully mixed state.

Finally, we consider the the mutual information \cite{wolf_area_2008, wilms_finite-temperature_2012,li_measurement-driven_2019},
\begin{equation}\label{mutual information}
    I_{AB} = S_\text{vN}(\rho_A) + S_\text{vN}(\rho_B) - S_\text{vN}(\rho_{A\cup B})\,
\end{equation}
which captures the total correlations between two disjoint subsystems $A$ and $B$. This quantity can be used to diagnose phase transitions at finite temperature \cite{wolf_area_2008, li_measurement-driven_2019, wilms_finite-temperature_2012}. Here, we can calculate it analytically in the normal phase by using Eq.~\eqref{split cov mat}, in combination with Eq.~\eqref{vN entropy}. To obtain the subsystem entropy, we construct the subsystem covariance matrices from Eq.~\eqref{split cov mat},
\begin{equation}
    \sigma^A = \sigma^B = \begin{pmatrix}
        X_{11} &  K_{11} \\ K_{11} & P_{11}\
    \end{pmatrix}  \,,
\end{equation}
where the matrix elements are given by Eq.~\eqref{split cov matrix elements}. The matrices are identical for the two subsystems as they are equal in size and 
due to the permutation symmetry of the model. 
The symplectic eigenvalues of these covariance matrices then read as
\begin{equation}
    \nu^A = \nu^B = \sqrt{1 + \frac{12 J^2}{\Gamma^2 - \Gamma_c^2}}\,,
\end{equation}
which are similar to, but distinct from, the symplectic eigenvalue of the total system covariance matrix $\nu$ given by Eq.~\eqref{total system symplectic eigenvalue}. Plugging these eigenvalues into \cref{vN entropy,mutual information}, we find the mutual information in the normal phase. Similar to the von Neumann entropy, the mutual information diverges logarithmically at the phase boundary as $I_{AB} \sim \frac{1}{2}\log \gamma$. Again invoking the critical scaling, this quantity also grows logarithmically with the system size as
\begin{equation}\label{MI cp}
    I_{AB} \sim \frac{1}{4}\log N\,,
\end{equation}
with the same coefficient as the von Neumann entropy. Interestingly, the mutual information shows same scaling as well as the universal coefficient at a thermal phase transition \cite{wilms_finite-temperature_2012}. This may be expected as this phase transition is shown to be effectively thermal despite the non-equilibrium dynamics \cite{paz_driven-dissipative_2021,paz_driven-dissipative_2022,paz_time-reversal_2021}.  However, as we discuss next, the thermal analogy ceases to hold into the ordered phase. 

\begin{figure}
    \hspace*{-0.5cm} 
    \includegraphics[scale=0.33]{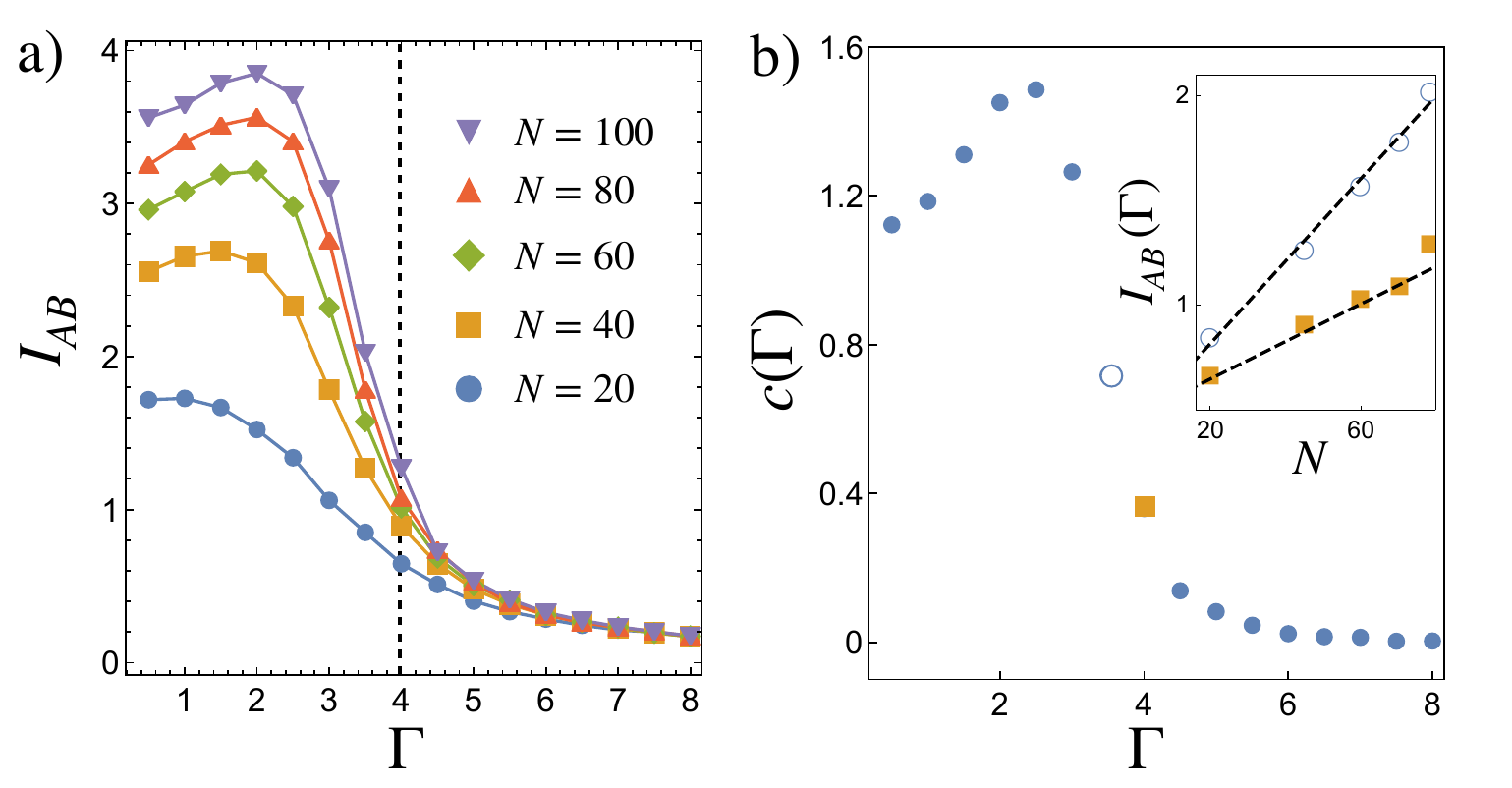}
     \caption{(a) Mutual information as a function of $\Gamma$ for various system sizes, with $J = 1, \Delta = 1$. This quantity scales logarithmically with system size not only at the critical point ($\Gamma_c =4$ marked by the dashed line) but also in the ordered phase. (b) The coefficient of the logarithmic dependence as a function of $\Gamma$. The inset explicitly shows the logarithmic fit at the critical point (squares) and a point inside the ordered phase (open circles).} 
     \label{fig MI}
\end{figure}

In Fig.~\ref{fig MI}(a), we plot the mutual information  for different system parameters as well as different system sizes. These plots are obtained numerically  using  quantum trajectories while incorporating the permutation symmetry based on Ref.~\cite{Zhang_2018}.
Approaching the critical point from the normal phase, the mutual information increases with system size. 
Specifically, it grows logarithmically at the phase boundary [see the squares in the inset of Fig.~\ref{fig MI}(b)], in harmony with Eq.~\eqref{MI cp}. 
We note a slight discrepancy in the numerically obtained coefficient $\sim 0.36$ of the logarithmic dependence against the theoretically predicted value of $1/4$; this could be attributed to finite-size effects and possibly the dynamical slowdown at the critical point requiring evolution up to times that grow with system size as $\sqrt{N}$ \cite{paz_driven-dissipative_2022}. Specifically, it is difficult to numerically access the late times necessary to ensure we have converged to the steady state, let alone the memory cost of storing the density matrix over longer time scales and for larger system sizes. With these considerations into account, we have performed the numerical simulations with a time step of $\delta t = .001$ over 2000 trajectories, and we have averaged over the last 100 time steps, skipping every 5, in the dynamics. At the critical point, the total evolution time is chosen as $t_f = 10\Gamma^{-1}\sqrt{N}$ to account for the finite-size scaling of the critical dynamics. In the ordered phase, we instead take $t_f = 10\Gamma^{-1}\log N$ (although for the systems sizes considered, the latter times scales are at most different roughly by a factor of 2). 

Surprisingly, Fig.~\ref{fig MI} shows that the mutual information does not peak at the phase transition and still grows with the system size in the ordered phase where $\Gamma < \Gamma_c = 4$ (for $\Delta=1$), and even exhibits a pronounced peak well within the ordered phase.
In fact, the mutual information appears to scale logarithmically with system size even in the ordered phase, $I_{AB}\sim c(\Gamma) \log N$, with a nontrivial coefficient $c(\Gamma)$; see Fig.~\ref{fig MI}(b).   
In contrast, the mutual information in thermal equilibrium diverges at the critical point while approaching a constant of log 2 deep in the ordered phase due to the Ising spontaneous symmetry breaking \cite{wilms_finite-temperature_2012}. The persistent critical scaling in the ordered phase of our non-equilibrium model hints at a hidden criticality that cannot be detected by two-point correlation functions. On the other hand, mutual information provides an upper bound on all correlations \cite{Wolf_2008} and does not ``overlook'' any correlations. Hidden critical behavior in the non-equilibrium setting of quench dynamics has been recently identified through measures such as mutual information as well \cite{Paul_2024}. 
While we do not fully understand the origin of this behavior in the present context, we speculate that it follows from the contribution of different sectors each characterized by a distinct value of the total angular momentum. 
Note that the permutation symmetry of the model leads to a mixture of different sectors in the steady state \cite{chase_collective_2008, baragiola_collective_2010}. It is thus possible that the interplay of the permutation symmetry and the ordering gives rise to the nontrivial behavior of the mutual information.

\section{Entanglement Measures}\label{sec:ent_measures}
In this section, we investigate three prominent measures of quantum entanglement: logarithmic negativity, quantum Fisher information, and spin squeezing. Importantly, they are all proper measures of entanglement even for mixed states, while they characterize different aspects of the entanglement. 
Using a mix of analytical and numerical methods, 
we calculate each of these quantities throughout the phase diagram.

\subsection{Logarithmic Negativity}\label{sec log neg}
As a first measure of quantum entanglement in a mixed state, we consider the logarithmic negativity \cite{vidal_computable_2002, plenio_logarithmic_2005}
\begin{equation}\label{log negativity}
    E_{\mathcal{N}} = \log_2 \tr (|\rho^{T_{B}}|_1)\,,
\end{equation}
 where $T_B$ denotes the partial transpose of a subsystem $B$, and $|\bullet|_1$ signifies the trace norm. The partial transpose only affects the coherences of the density matrix, which could violate its positivity and lead to negative eigenvalues. The logarithmic negativity thus captures the degree to which positivity is violated due to the entanglement of the subsystems A and B, and is in fact an entanglement monotone \cite{plenio_logarithmic_2005}. This quantity can also be used to detect phase transitions and critical phenomena in many-body systems \cite{wichterich_universality_2010, lu_singularity_2019}.\par

 Equation \eqref{log negativity} requires access to the full density matrix in order to compute the singular values of its partial transpose. However, for Gaussian states, we can calculate this quantity using the covariance matrix formalism. Specifically, we need the covariance matrix for a system split in two halves. In terms of the symplectic eigenvalues of the covariance matrix, the logarithmic negativity can be then computed as \cite{wichterich_universality_2010, adesso_extremal_2004}
 \begin{equation}\label{log neg cov mat}
     E_{\mathcal{N}} = -\sum_i \log_2\left(\text{min}(\tilde{\nu}_i, 1) \right)\,,
 \end{equation}
 where $\tilde{\nu}_i$s are the symplectic eigenvalues upon partial transposition of the density matrix, which is equivalent to sending $p_B \to - p_B$ in the covariance matrix \cite{wichterich_universality_2010, adesso_extremal_2004}. The violation of positivity in the density matrix is equivalent to the violation of the bound $\nu \geq 1$. Computing these new symplectic eigenvalues, we find only one that satisfies $\tilde{\nu} < 1$,
 \begin{equation}
     \tilde{\nu} = \sqrt{1 + \frac{4J ( 4J - \sqrt{\Gamma^2 + 16 (J - \Delta)^2}) }{\Gamma^2 - 16\Delta(2J-\Delta)}}\,.
 \end{equation}
 Plugging this into Eq.~\eqref{log neg cov mat}, we find that $E_\mathcal{N}$ is indeed \textit{finite} throughout the normal phase. Specifically upon approaching the phase transition, $\Gamma \to \Gamma_c$, the symplectic eigenvalue becomes $\tilde{\nu} \to 1/\sqrt{2}$, resulting in the logarithmic negativity $E_\mathcal{N} = \log_2\sqrt{2} = 1/2$ everywhere along the phase boundary. This behavior is also shared by other measures of entanglement (see below), and is reminiscent of finding the same effective temperature at the phase boundary \cite{paz_driven-dissipative_2021,paz_driven-dissipative_2022}. We can conclude that while quantum correlations do not govern the phase transition, they are still present. In contrast, divergent quantum fluctuations emerge at a critical quantum ground state \cite{wichterich_universality_2010}. 

 \begin{figure}
      \hspace*{-0.5cm}
     \includegraphics[scale=0.37]{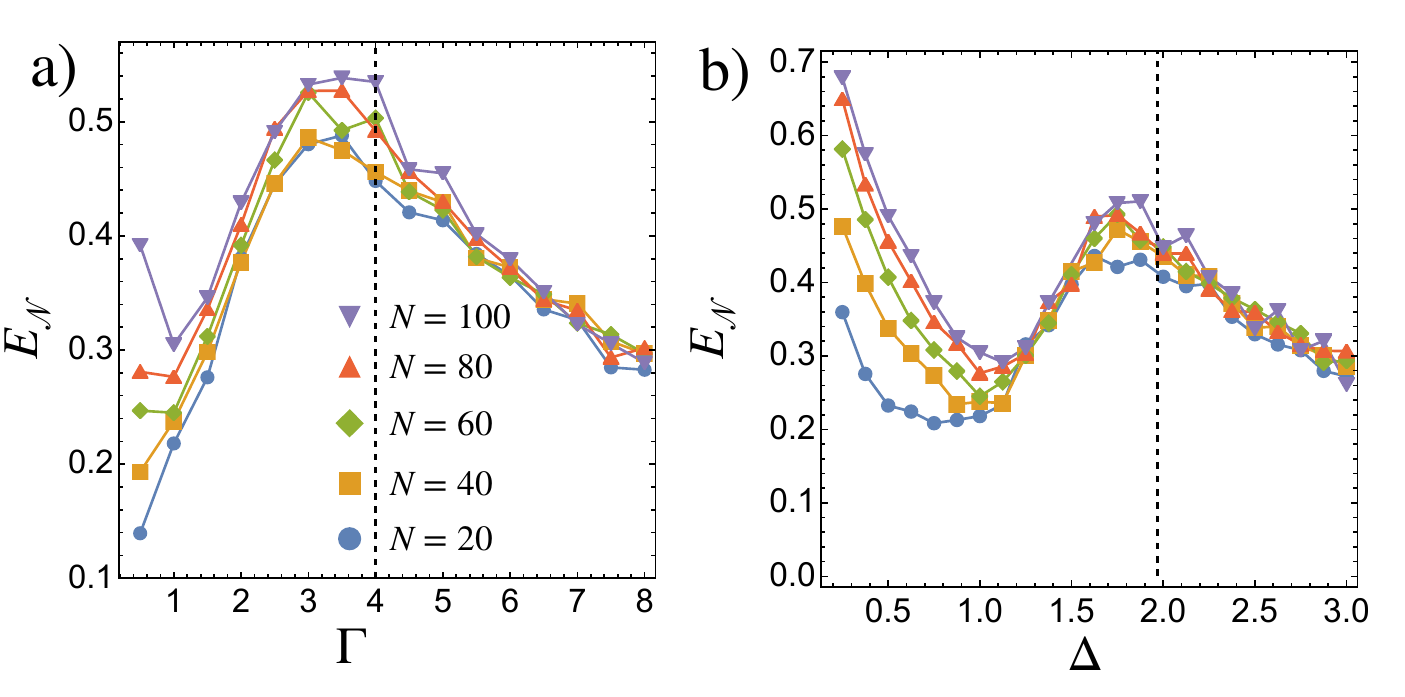}
      \caption{Logarithmic negativity $E_{\mathcal{N}}$ versus (a) $\Gamma$ with $J = \Delta = 1$ and (b) $\Delta$ with $J = \Gamma = 1$. In both cases, $E_\mathcal{N}$ peaks close to the critical point (marked by the dashed lines) and whose value is consistent with $E_\mathcal{N} = 1/2$. A slight deviation from the critical point is due to finite-size effects. Furthermore, the peak exhibits a kink. The features at small $\Gamma$ in (a) could be due to the limited evolution time. The growing logarithmic negativity at small $\Delta$ in (b) is likely an artifact of quantum trajectories; see the text.}
    \label{fig LN}
\end{figure}

 In the ordered phase, we rely on quantum trajectories to calculate the logarithmic negativity. In  Fig.~\ref{fig LN}(a), we plot the logarithmic negativity as a function of $\Gamma$ at fixed $\Delta=1$. We first observe that the peak appears close to the phase boundary, and is consistent with the analytical prediction, $E_\mathcal{N} = 1/2$, which has been obtained in the thermodynamic limit.  We note that the peak location is slightly shifted away from the critical point ($\Gamma_c = 4$ at $\Delta=1$) which is likely due to finite-size effects as the peak roughly converges toward the critical point as the system size increases. 
 We also notice that the logarithmic negativity in Fig.~\ref{fig LN}(a) grows with system size when $\Gamma \ll J$; however, this is likely due to the limitations of quantum trajectories. Small $\Gamma$ requires longer evolution times and storage of a longer density matrix history, which become inaccessible in this regime. We have adopted the  total evolution time of $t_f = 10 \Gamma^{-1} \log N$ in the ordered phase, but convergence at small $\Gamma$ might require longer times. 
 
  In  Fig.~\ref{fig LN}(b), we  plot the logarithmic negativity as a function of $\Delta$ at a fixed $\Gamma=1$. Notice that there are two critical points one at small $\Delta$ and another at ${\Delta_c =1.97}$. Again, we find that the logarithmic negativity peaks roughly at 1/2 and slightly away from the critical near $\Delta =2$. Furthermore, the logarithmic negativity in both panels exhibits a kink as one enters the ordered phase. Such kinks have been observed at thermal phase transitions \cite{lu_singularity_2019} as well as the open Dicke model \cite{boneberg_quantum_2021}. We also notice that the logarithmic negativity grows as we approach  the phase boundary near $\Delta = 0$. In fact, it appears to overshoot the theoretical prediction and shows a pronounced system size dependence. These features however are likely an artifact of quantum trajectories. 
  Indeed, the model with $\Delta=0$ can be solved exactly and is shown not to support correlations spreading \cite{foss-feig_solvable_2017}. In contrast, the non-Hermitian Hamiltonian, ${H_{\rm NH} = H  -i\sum_i L_i^\dagger L_i =H - i \frac{\Gamma}{4} \sum_i\sigma^z_i}+$ const, employed in quantum trajectories is an Ising model with an imaginary transverse field (at $\Delta=0$), and allows correlations to propagate.  The fact that the jump terms and the non-Hermitian Hamiltonian are treated on different footings in quantum trajectories should be responsible for the unusual behavior near $\Delta=0$. Convergence  may require a large number of quantum trajectories that is not accessible in our numerics.

\subsection{Quantum Fisher Information}\label{sec QFI}
The quantum Fisher information $F$, although typically used in quantum metrology, is a useful measure of entanglement \cite{ma_fisher_2009, hyllus_fisher_2012, li_entanglement_2013, hauke_measuring_2016, toth_multipartite_2012}. This quantity bounds the precision one can attain when performing a phase estimation measurement corresponding to the transformation $U = \exp(i \theta O)$ with the phase $\theta$ and operator $O$. To saturate this bound, the system needs to be entangled \cite{giovannetti_advances_2011, ma_fisher_2009, hyllus_fisher_2012, li_entanglement_2013, hauke_measuring_2016, toth_multipartite_2012}. For a spin system, it has been shown that the quantum Fisher information density corresponding to the total spin operator $\frac{1}{2}S_{\mathbf{n}} = \frac{1}{2}
\sum_i \mathbf{n} \cdot \boldsymbol{\sigma}_i$  pointing along the unit vector $\mathbf{n} = (n_x, n_y, n_z)$ (using the notation $\boldsymbol{\sigma}_i = (\sigma^x_i, \sigma^y_i, \sigma^z_i)$), can indicate whether or not a state is $k$-particle entangled \cite{ma_fisher_2009, hyllus_fisher_2012, li_entanglement_2013, hauke_measuring_2016, toth_multipartite_2012}. The overall factor of $1/2$ is introduced so that the spin operators have a spectrum of unit width per particle. Specifically, a state of $N$ spin-$\frac{1}{2}$ particles is at least $(m+1)$-particle entangled if
\begin{equation}\label{QFI bound}
    \frac{F}{N} = f > m\,,
\end{equation}
for $m$ a divisor of $N$. This bound provides a direct way to determine the presence of entanglement from the quantum Fisher information density.
More precisely, the quantum Fisher information corresponding to a density matrix $\rho$ and a given operator $O$ is defined as
\begin{equation}\label{QFI def}
    F(\rho, O) = 2 \sum_{i,j} \frac{(\lambda_i - \lambda_j)^2}{\lambda_i + \lambda_j}|\bra{i}O\ket{j}|^2\,,
\end{equation}
where $\lambda_i$ and $\ket{i}$ denote the eigenvalues  and eigenvectors  of the density matrix, respectively. However, these quantities are difficult to obtain analytically. Again,  Gaussian states permit an analytical expression of the quantum Fisher information in terms of the covariance matrix and the displacement vector \cite{safranek_optimal_2016}.
\begin{figure}
  \centering
    \includegraphics[width=\linewidth]{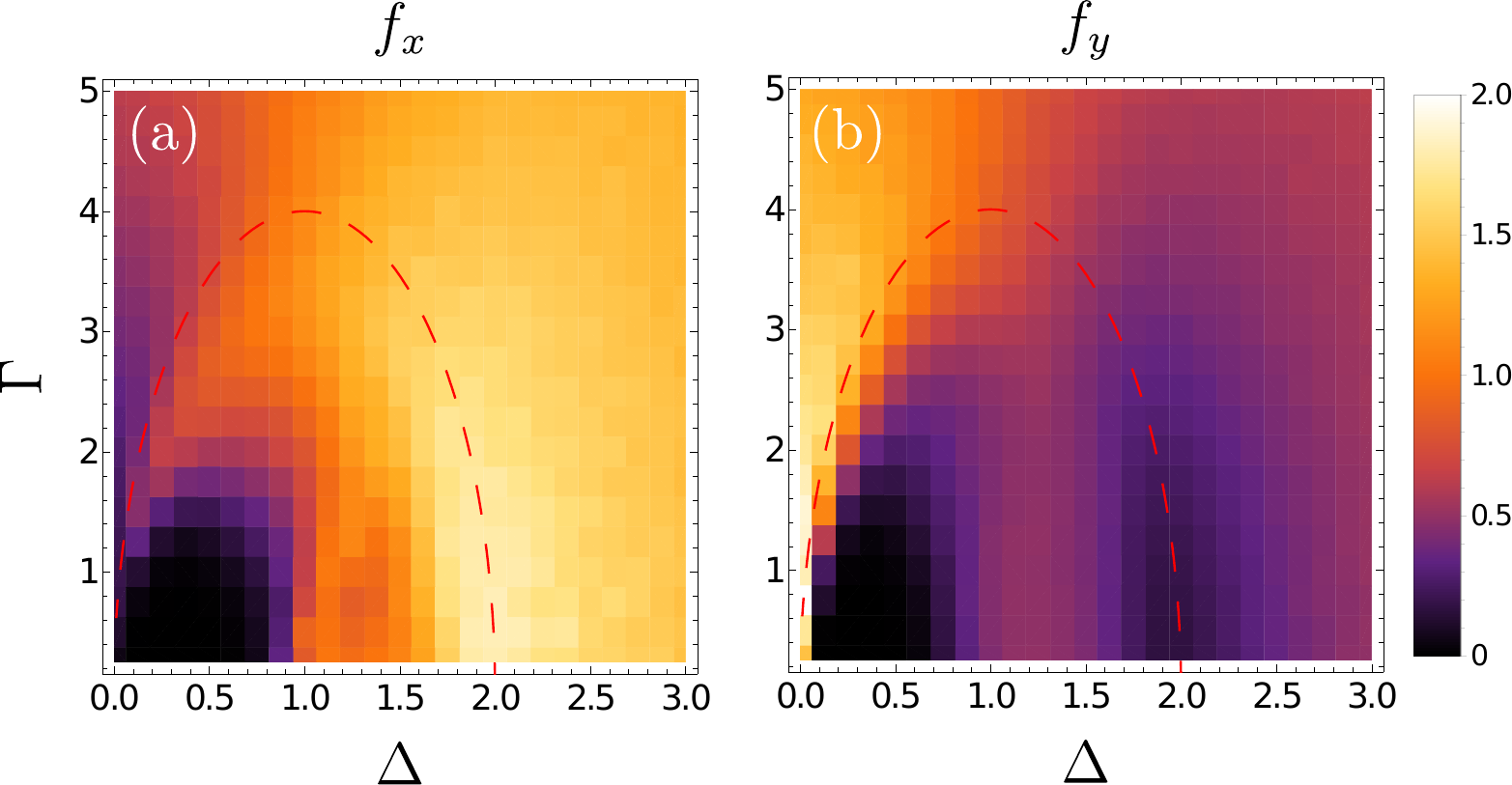}
    \caption{Quantum Fisher information densities $f_{x,y}$ corresponding to the operators $S_{x,y}/2$ for $N=100$ spins. (a) $f_x$ is maximal in the limit $\Delta \to 2, \Gamma \to 0$, approaching the theoretically predicted value of $f_x = 2$. (b) $f_y$ approaches its theoretically predicted maximum value of $f_y = 2$ in the limit $\Delta \to 0, \Gamma \to 0$. In both cases, the quantum Fisher information is vanishing in the corner of the ordered phase where $\Delta, \Gamma \to 0$. In these plots, $J=1$.} 
    \label{fig_fx_fy}
\end{figure}
For a generator of the transformation that is a collective spin operator in the $x-y$ plane,  we can write the unitary in the form 
\begin{equation}
    U(\theta) = e^{i \mathbf{r}^T \Omega \boldsymbol{\gamma}(\theta)} \,,
\end{equation}
where $\boldsymbol{\gamma}(\theta) = \sqrt{\frac{N}{2}}\theta\times(n_y, n_x)$; we recall that $\frac{1}{2}S_x = \sqrt{N/2}x$, $\frac{1}{2}S_y = -\sqrt{N/2}p$, and $\mathbf{r} = (x, p)$. The above unitary transformation merely shifts the displacement vector,
\begin{equation}
    \mathbf{d}(\theta) = \mathbf{d} + \boldsymbol{\gamma}(\theta)\,.
\end{equation}
For a transformation of this type, linear in the canonical operators, $F$ is given by \cite{safranek_optimal_2016}
\begin{equation}\label{Gaussian QFI Shift}
    F = \dot{\boldsymbol{\gamma}}^T(\theta) \sigma^{-1} \dot{\boldsymbol{\gamma}}(\theta)\,,
\end{equation}
where the dot denotes a derivative with respect to $\theta$ (in this case resulting in a fixed vector). 
Substituting the covariance matrix in Eq.~\eqref{covariance mat} into Eq.~\eqref{Gaussian QFI Shift}, we find the quantum Fisher information densities
\begin{align}
    f_x &= \frac{\Gamma^2 -16\Delta(J - \Delta)}{\Gamma^2 + 16(J-\Delta)^2}\,,\\
    f_y &= 1 + \frac{16 J (J - \Delta)}{\Gamma^2 + 16(J-\Delta)^2}\,,
\end{align}
corresponding to the collective operators $S_x$ and $S_y$, respectively,
everywhere in the normal phase. Interestingly, each of these quantities is bounded from above by $f_x,f_y \leq 2$. In fact, $f_x + f_y = 2$ is an exact relation, even at the phase boundary. Furthermore, through the bound given by Eq. \eqref{QFI bound}, we can conclude that the system is at least 2-particle entangled. The quantum Fisher information density along the $z$-direction is zero within the normal phase in the thermodynamic limit, as the collective spin is fully polarized, thus no fluctuations, in this direction.

\begin{figure}
    \centering
    \includegraphics[width=\linewidth]{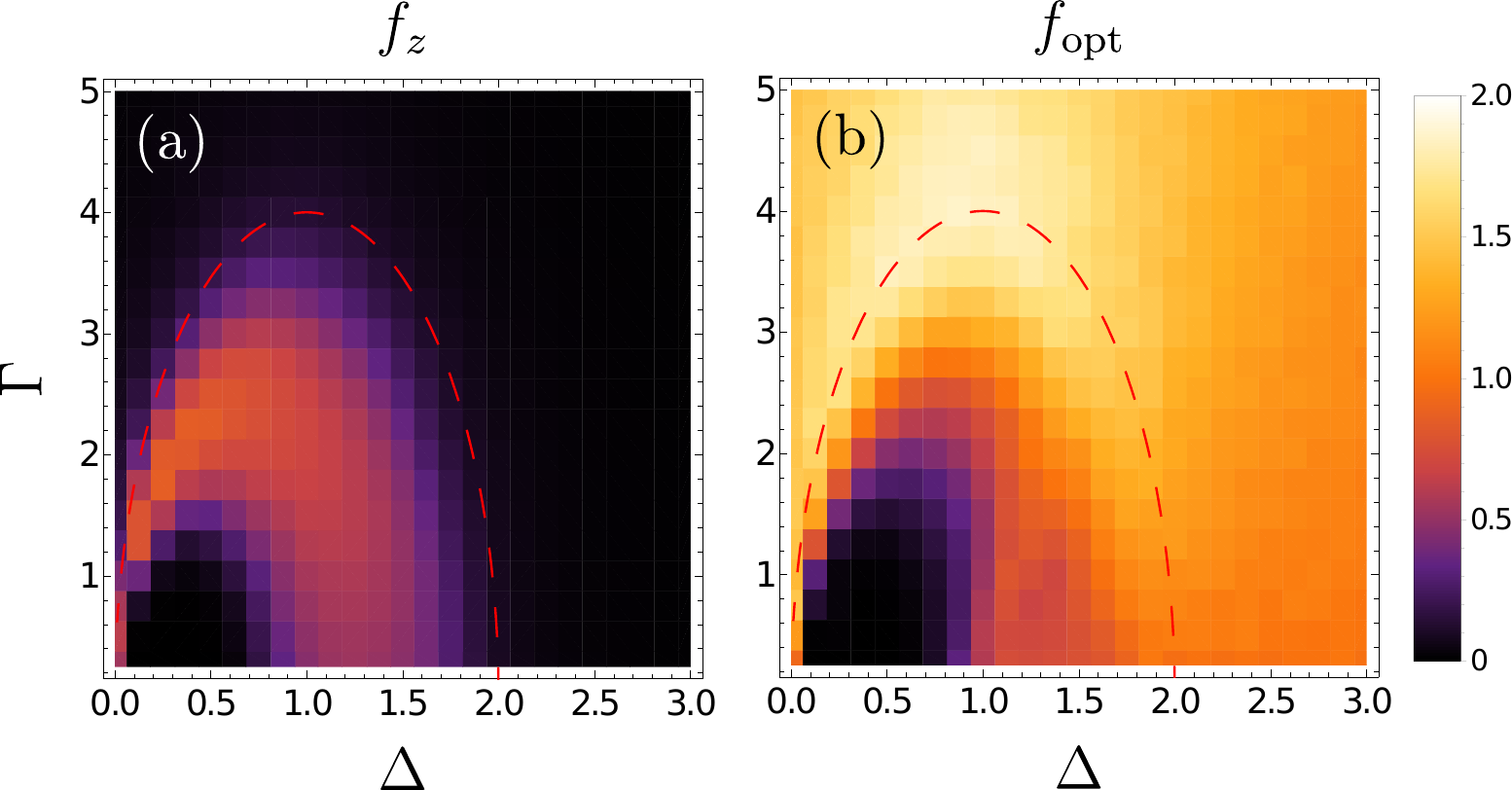}
    \caption{Quantum Fisher information density $f_z$ corresponding to the operator $S_z/2$ as well as the optimal density $f_{\rm opt}$ for $N=100$ spins. (a) $f_z$ is nonzero in the ordered phase where the collective spin is no longer fully polarized along the $z$ direction, allowing for nontrivial fluctuations of $S_z$. (b) $f_{\rm opt}$ peaks at  the phase boundary and is consistent with the analytical prediction of $f_{\text{opt}} = 2$. The Fisher information densities are all vanishing in a dead zone in the corner of the ordered phase where $\Delta, \Gamma \to 0$.}
    \label{fig fz fopt}
\end{figure}

\begin{figure}
    \centering
    \includegraphics[width=\linewidth]{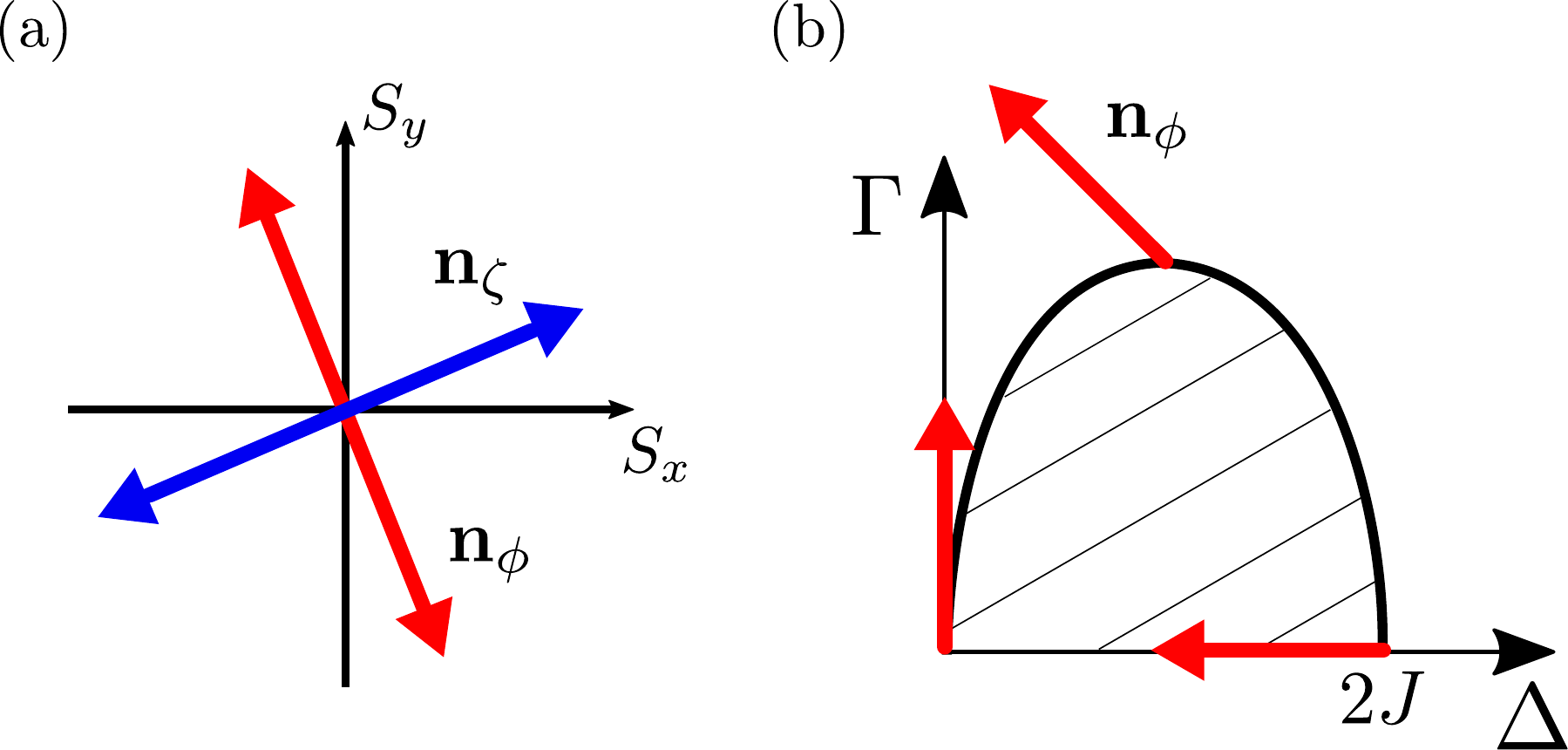}
    \caption{(a) A schematic diagram depicting the direction of the gapless ($\mathbf{n}_\phi$) and gapped ($\mathbf{n}_\zeta$) modes in the $S_x$-$S_y$ plane. (b) The depiction of the gapless mode along the phase boundary. The corresponding vector $\mathbf{n}_\phi$ rotates from vertical to horizontal as the phase boundary is traversed from left to right. }
    \label{modes fig}
\end{figure}

The above results hint towards an optimal direction $\mathbf{n}$ along which $F$ is maximized. This optimal direction can be determined by solving for the $n_x, n_y$ (defining a unit vector while setting $n_z=0$) that maximizes $F$, and is given by
\begin{align}\label{optimal directions}
    n_x^* = -\sqrt{\frac{1}{2} - \frac{2(J-\Delta)}{\sqrt{\Gamma^2 + 16(J-\Delta)^2}}},
\end{align}
and $n_y^* = \sqrt{1-{n_x^*}^2}$ up to an overall sign. Plugging these expressions into Eq. \eqref{Gaussian QFI Shift}, we find the optimal quantum Fisher information density one can achieve in the normal phase,
\begin{equation}\label{eq:optimal_F}
    f_{\rm opt} = 1 + \frac{4J}{\sqrt{\Gamma^2 + 16(J-\Delta)^2}}\,.
\end{equation}
We see then see that $1 \leq f_{\rm opt} \leq 2$ throughout the normal phase, again suggesting that the steady state is at least 2-particle entangled in this phase. Notice that the upper bound is saturated at the phase transition.  The optimal direction (dropping the $n_z$ component for ease of notation) on the phase boundary changes continuously from $\mathbf n =(1,0)$ at the rightmost critical point ($\Delta=2J, \Gamma\to 0$) to  $\mathbf n =(0,1)$ at the leftmost critical point ($\Gamma,\Delta/\Gamma\to 0$). 
At criticality, the optimal direction coincides exactly with the ``gapless mode" of the system (i.e., the critical mode),  given by  $\mathbf{n}_\phi = (\sqrt{\Delta}, -\sqrt{2J-\Delta})/\sqrt{2J}$, as shown in a previous work by some of the authors \cite{paz_time-reversal_2021}; see also \cref{modes fig}. Using this notation, the quantum Fisher information density is $f_\phi  =2$ everywhere on the phase boundary. Along the ``gapped" direction, $\mathbf{n}_\zeta = (\sqrt{2J-\Delta}, \sqrt{\Delta} )/\sqrt{2J}$, we instead find $f_\zeta = 0$. These results are intuitive as the quantum Fisher information, a useful measure of quantum metrology, is  sensitive to fluctuations \cite{giovannetti_quantum_2006, giovannetti_advances_2011}; it is precisely the gapless mode that exhibits the largest (in fact, divergent) fluctuations at criticality, while the gapped mode has negligible fluctuations, resulting in a vanishing quantum Fisher information density. 
Lastly, we remark that a saturation of the bound in Eq.~\eqref{QFI bound} implies that the state is an $m$-particle GHZ state \cite{hyllus_fisher_2012}. Interestingly, the optimal quantum Fisher information in Eq. \eqref{eq:optimal_F} saturates this bound at the phase boundary with $m =2$, hinting towards the emergence of a 2-particle GHZ-like state at the phase transition. The precise nature of this state is not clear to us, which we leave to future work. \par

To investigate the ordered phase, we once again employ quantum trajectories and consider a system of $N=100$ spins. The density plots for $f_x, f_y$ are given in \cref{fig_fx_fy}. The maximum of $f_x$ ($f_y$) is close to the phase boundary on the right (left) side of the phase diagram. This is also consistent with the optimal direction being parallel to $x$ ($y$) in this limit. 
Figure~\ref{fig fz fopt}(a) shows that $f_z$ also assumes a finite value in the ordered phase, as the spins are no longer fully polarized, hence allowing for fluctuations, along the $z$ direction. 
Finally, we plot the optimal quantum Fisher information density in \cref{fig fz fopt}(b) which clearly peaks at the phase boundary. 
Notice that all the above quantities feature a dead zone in the ordered phase close to $\Gamma, \Delta \to 0$ where the state becomes increasingly mixed; see Eq.~(\ref{purity}). 
The quantum trajectory simulations here are performed by evolving up to $t_f = 10 \Gamma^{-1}$ with the time step $\delta t = 0.1$ and averaging over 1000 trajectories. For each data point, the quantum Fisher information is also averaged over the last ten time steps of the dynamics.

 \subsection{Spin Squeezing}\label{sec squeezing}
 Squeezed states are studied extensively, in part due to their applications to quantum metrology \cite{toth_quantum_2014, gross_spin_2012, ma_quantum_2011-1}. Technically, a state is squeezed when the variance of one or more quadratures is less than that of symmetric states such as coherent states \cite{rosas-ortiz_coherent_2019}. 
 For spin operators, this property can be quantified by the Wineland squeezing parameter \cite{Wineland_1992,Pezze_2018}
 \begin{equation}\label{spin squeezing parameter}
     \xi = \frac{\text{min}(\Delta S_{\mathbf{s}_\perp}^2)}{s^2 N}\,,
 \end{equation}
 where $s$ is the magnitude of the magnetization vector per particle $\mathbf s$, $\Delta O^2 = \langle O^2\rangle - \langle O \rangle^2$ denotes the variance of operator $O$, and $\mathbf{s}_\perp$ defines a unit vector perpendicular to $\mathbf{s}$. The minimization is performed over all directions perpendicular to $\mathbf{s}$. While for a coherent state,  $\xi = 1$, a squeezed state state will correspond to $\xi < 1$. The squeezing parameter is related to the quantum Fisher information as they are both sensitive to fluctuations and both indicate if the state is entangled \cite{toth_quantum_2014, ma_quantum_2011-1}. In fact, the squeezing parameter is directly related to the concurrence, and $\xi < 1$ implies that the state is not only squeezed, but is also entangled \cite{ma_quantum_2011-1}.

 Using the method described in Appendix \ref{ordered phase appendix}, we can calculate all spin correlations including those in the ordered phase. For convenience, we define the $z$-direction along $\mathbf{s}$, write the perpendicular spin as $S_{\mathbf{s}_\perp} = \tilde{\mathbf{S}}\cdot \mathbf{s}_\perp = \cos(\phi) \tilde{S}_x + \sin(\phi) \tilde{S}_y$ and minimize Eq. \eqref{spin squeezing parameter} with respect to $\phi$. The tilde indicates that the spin components are defined in a rotated frame such that $\langle \tilde{S}_z \rangle = Ns$. 
 
 In the normal phase, the covariance matrix in Eq. \eqref{covariance mat} already provides us with the necessary ingredients to write Eq. \eqref{spin squeezing parameter} as 
 \begin{equation}\label{xi to minimize}
     \xi =
     \text{min}_\phi(\sigma_{11} \cos^2 \phi  + \sigma_{22} \sin^2 \phi  + 2 \sigma_{12} \cos \phi \sin \phi )\,.
 \end{equation}
 We have used the fact that $\tilde{S}_{x} = -S_{x}, \tilde{S}_y = S_y$ in the normal phase\footnote{The negative sign is due to a 180$^\circ$ rotation around the $y$ axis to bring the fully polarized state from the negative to the positive $z$ direction.}, and identified $x = S_x/\sqrt{2N}$ and $p = -S_y/\sqrt{2N}$.
 We note that the above expression is simply the minimum eigenvalue of the $\sigma$ matrix which is given by
 \begin{equation}
     \xi = \frac{1}{1+4J / \sqrt{\Gamma^2+16(J-\Delta)^2}}\,.
 \end{equation}
 Interestingly, we find $\xi = 1/2$ exactly all along the phase boundary. This value is a lower bound on the squeezing parameter all throughout the phase diagram (see the discussion of the ordered phase below). These results are consistent with the 3 dB limit of squeezing at the onset of instability of parametric amplifiers \cite{MILBURN1981401}. From this perspective, the driven-dissipative phase transition of the infinite-range Ising model is qualitatively similar to the threshold instability of a parametric amplifier. 
 Finally, at the phase boundary, we solve for $\phi$ that minimizes the expression in \cref{xi to minimize} and find the unit vector  $\mathbf{s}_\perp^* = (\sqrt{2J-\Delta},  -\sqrt{\Delta})/\sqrt{2J}$, which coincides, upon the sign conversion, with the direction of the gapped mode $\mathbf{n}_\zeta =  (\sqrt{2J-\Delta}, \sqrt{\Delta} )/\sqrt{2J}$ in the original frame.

 In the ordered phase, while we do not have analytical expressions,  we can numerically compute the correlation functions as detailed in Appendix \ref{ordered phase appendix}. In Fig.~\ref{fig squeezing}, we present the  density plot of the squeezing parameter in the phase diagram. One can see that the  state is most squeezed at the phase boundary. Furthermore, the steady state is not squeezed ($\xi \geq 1$) deep inside the ordered phase close to $\Delta, \Gamma =0$. This should be expected as the state becomes increasingly mixed in this region, previously referred to as the dead zone in \cref{sec QFI}. 
 
 \begin{figure}
    \centering
    \includegraphics[scale=.35]{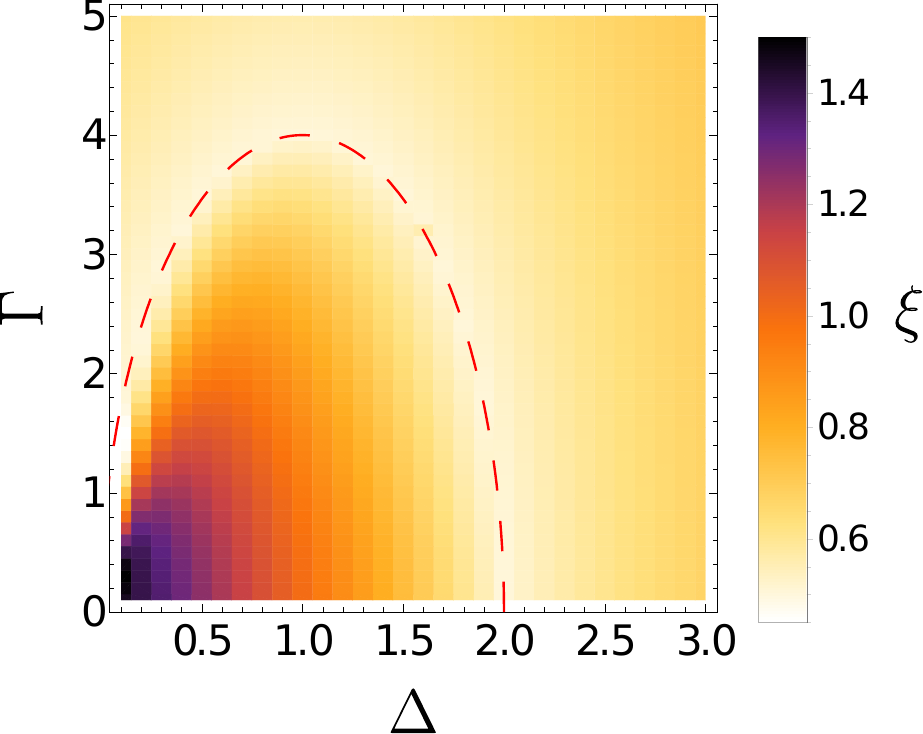}
    \caption{Squeezing parameter $\xi$ over the phase space in the phase diagram; we have set $J=1$. The steady state is most squeezed along the phase boundary with $\xi=1/2$. In the corner region within the ordered phase, the steady state is not squeezed as it becomes increasingly mixed.}
    \label{fig squeezing}
\end{figure}

\section{Long-range Ising Model}\label{sec_long_range_Ising}
In this section, we finally turn to the DDIM with long-range interactions, $0\le\alpha\le 3$; we also include $\alpha=0$ for benchmarking. 
Here, we use the machinery of matrix product states (MPS), specifically the methods developed in \cite{Mascarenhas_2015,Kamar_splitting_2024}, allowing us to simulate systems up to $N=50$ spins. While computing correlation functions (being linear in the density matrix) are straightforward, computing entanglement and information measures (generally nonlinear in the density matrix) are rather challenging. For this reason, we focus on correlation functions and compute only the squeezing parameter, given its relatively simple form in terms of correlators of collective operators. 
We consider a  1D array of $N$ spin-1/2 particles  in a transverse field $\Delta$, 
and subject to long-range Ising interactions with the exponent $\alpha$, cf. the Hamiltonian in \cref{eq:full_Hamiltonian}. We also consider dissipative dynamics due to the individual atomic decay at the rate $\Gamma$.

\begin{figure*}
    \centering
    \includegraphics[width=1.05\linewidth]{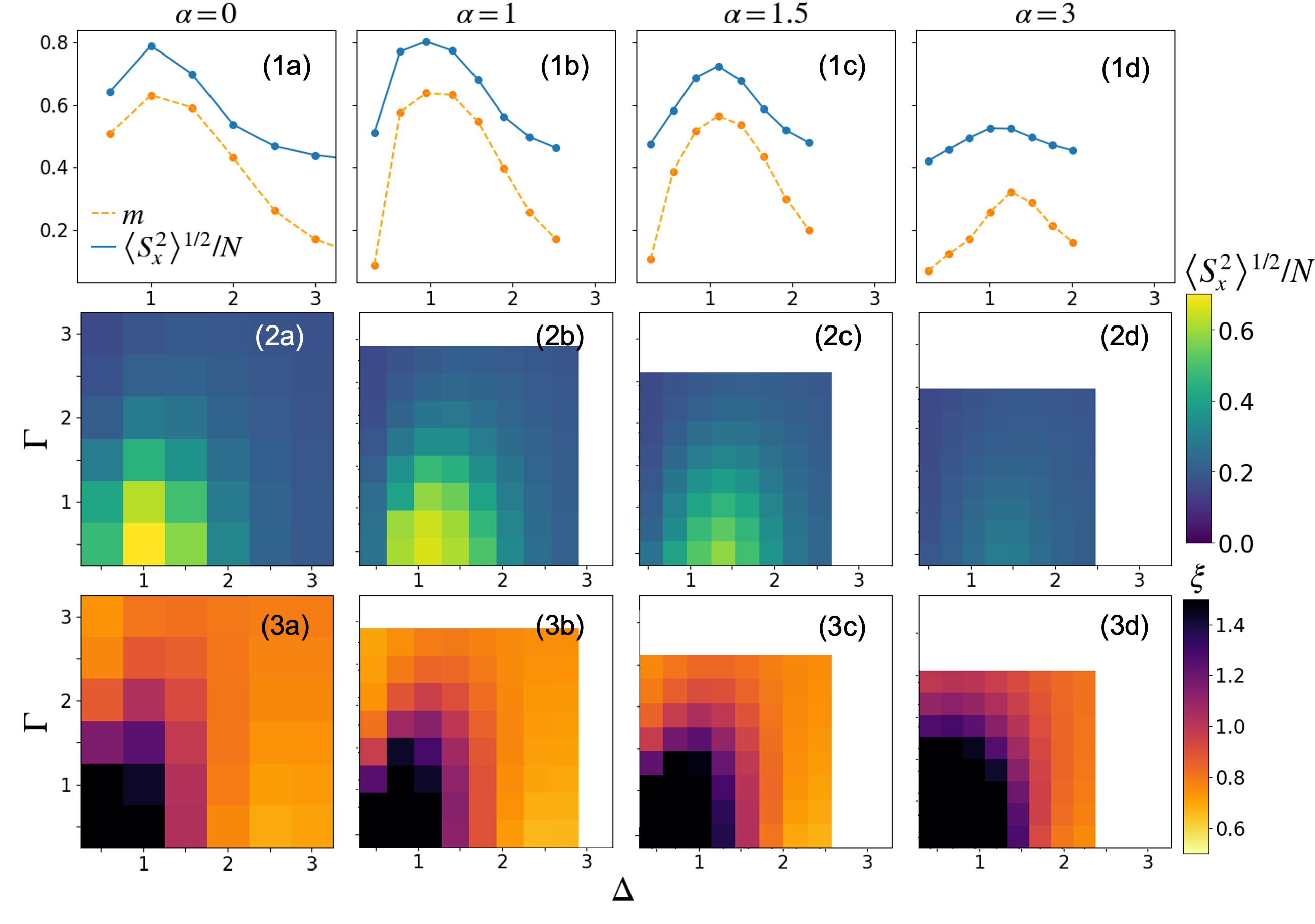}
    \caption{Order parameter, collective spin correlators, and squeezing in the open long-range Ising model with $N=50$ spins. (1) The order parameter $m$ and collective spin correlators $\langle S_x^2\rangle ^{1/2}/N$ follow similar trends. The order parameter $m=\langle S_x\rangle/N$ is shown as a function of $\Delta$ for a fixed $\Gamma =1, \cdots ,0.5$ for a given $\alpha=0, \cdots, 3$ in the presence of a symmetry breaking $h=0.1,\cdots, 0.05$, respectively. The correlator $\langle S_x^2\rangle ^{1/2}/N$ is computed in the absence of a symmetry breaking field. (2) Density plot of collective operators in the absence of a symmetry breaking field. These plots provide a qualitative depiction of the phase diagram. Specifically, an ordered phase or a phase transition is not expected for $\alpha=3$.  (3) Wineland squeezing  parameter $\xi$ in the presence of a symmetry breaking field [same as (1)]. The overall pattern is qualitatively similar for different values of $\alpha$ although the dark region (dead zone) is larger for $\alpha>1$. 
    While for smaller values of $\alpha$, the maximum squeezing follows a similar qualitative pattern as the phase diagram, for the largest value of  $\alpha=3$, there is no such relation.}
    \label{fig:enter-label}
\end{figure*}

Using the MPS numerical simulation, we compute the expectation value $\langle S_a\rangle$ as well as $\langle S_a S_b\rangle$ where $a,b \in \{x,y,z\}$. In the ordered phase, the expectation values $\langle S_a\rangle$ for $a\in\{x,y\}$ should take a nonzero value reflecting the symmetry breaking phase transition. Specifically, we  define the order parameter $m=\langle S_x\rangle/N$; if nonzero, this parameter indicates symmetry breaking. 
However, such symmetry breaking does not always occur in the numerics due to the finite system size (interestingly though we observe a numerical symmetry breaking transition deep in the ordered phase and for smaller values of $\alpha$). Here, we do not aim to identify the exact phase transition, which would require a careful analysis, for example by examining the Binder cumulant and its scaling behavior. Rather, we focus on qualitative features of phase transitions (if any). To this end, we take two different approaches. The first approach is simply to add a small symmetry breaking field $h$ along the $x$ direction to give a slight bias. If a small $h$ leads to a large order parameter, we may conclude that we are in the ordered phase. In contrast, in the normal phase $m\sim h$ due to the finite magnetic susceptibility. 
In a second approach, we define the quantity $\langle S_x^2\rangle^{1/2}/N$ as a proxy for the order parameter $m$. 
In the normal phase where $m=0$, this quantity scales as $1/\sqrt{N}$ and becomes increasingly smaller for larger system sizes. In the ordered phase and with an explicit symmetry breaking, it  becomes identical to the order parameter for very large system sizes.
More importantly, even without explicit symmetry breaking ($m_x=0$), this proxy should be comparable to 1 deep in the ordered phase. 

In \cref{fig:enter-label}(1a-d), we compare and contrast these two quantities. The order parameter is computed with a small symmetry-breaking field (ranging from $h_x=0.1$ to 0.05 for $\alpha =0$ upto 3) while $\langle S_x^2\rangle^{1/2}/N$ is computed with no bias; here, we have plotted these quantities as a function of $\Delta$ for fixed $\Gamma$ and a given value of $\alpha$. The correlators are always larger than the order parameter, yet the trends are quite consistent. Furthermore, we depict the density plot of $\langle S_x^2\rangle^{1/2}/N$ in \cref{fig:enter-label}(2a-d) for different values of $\Gamma,\Delta$.
For $\alpha=0,1$ both  describing mean-field models, the  phase diagram should be identical although finite-size fluctuations are stronger for $\alpha=1$. Furthermore, $\alpha=1.5$ appears to follow a similar trend, although the ordered phase is expected to shrink due to fluctuations. In contrast, for  $\alpha=3$, a phase transition seems unlikely as the plotted quantities are quite smaller, and there is less variation with the system parameters. Indeed, this is consistent with the expectation that, akin to thermal phase transitions \cite{Dyson_1969}, a driven-dissipation phase transition is lacking for $\alpha>2$. In a closely related model \cite{sierant_dissipative_2021}, it was shown that no phase transition occurs for $\alpha>1.3$. While the susceptibility exhibits a pronounced peak, it should be attributed to the sensitivity to the quantum phase transition in the ground state \cite{ali2024signatures}. These figures provide a qualitative understanding of the phase diagram for different values of $\alpha$.

Next, we compute the Wineland squeezing parameter; see \cref{spin squeezing parameter}. To this end, we add a small symmetry breaking field, and compute both $\mathbf s = \langle \mathbf S\rangle/N$ as well as $C_{ab}\equiv\langle S_a S_b\rangle$. The former quantity gives the expectation value of the spin, while the latter quantity allows us to compute the minimum variance in a direction perpendicular to ${\mathbf S}$. This variance can be easily computed by defining the projection matrix $\Pi = I - {\mathbf n} \otimes  {\mathbf n}$ where we have defined the $3\times3$ identity matrix $I$ and the unit vector ${\mathbf n} = \mathbf s/ s$. Projecting out the axis parallel to $\mathbf s$ (or $\mathbf n$), the quantity $\text{min}(\Delta S_{\mathbf{s}_\perp}^2)$ is just given by the smallest nonzero eigenvalue of matrix product $\Pi C \Pi$.  
The squeezing parameter can be then computed from in a straightforward fashion. The results are shown in \cref{fig:enter-label}(3a-d). Going from $\alpha=0$ to larger values of $\alpha$, the overall pattern of squeezing does not alter significantly with the exception of the dead zone of entanglement which is increasing in size for $\alpha>1$. On the other hand, notice that the ordered phase should shrink (and not extend)  in the same regime, thus featuring the opposite tendency. We also observe that the minimum value of the squeezing parameter, while not saturating at 0.5 possibly due to finite-size effects, comes close to this value. Interestingly, this feature is not specific to smaller values of $\alpha$ and emerges even for $\alpha=3$ where no phase transition is expected. Therefore, the connection between the phase transition and the squeezing (or rather entanglement) features does not hold for larger values of $\alpha$. It is nevertheless interesting that all the models considered lead to some degree of entanglement close to the 3 dB limit reminiscent of the threshold behavior even if there is no phase transition.

 \section{Conclusion}\label{sec:conclusion}
 In this work, we have studied entanglement and information in driven-dissipative Ising models, specifically in relation to the disorder-to-order phase transitions in these systems. For the infinite-range Ising model, we have calculated the logarithmic negativity, quantum Fisher information, and spin squeezing  which constitute proper measures of entanglement for mixed states. We have provided analytical results in the normal phase, and have utilized quantum trajectories for relatively large system sizes (up to $N=100$ spins) to probe the ordered phase. We find consistent features where the entanglement peaks, features a kink, and takes a universal value at the phase transition. Furthermore, we have uncovered a connection between the gapless/gapped modes of the phase transition and the optimal Fisher information/squeezing.
 We have also computed mutual information that captures total (classical and quantum) correlations, and find that it scales logarithmically with system size not only at the phase transition but, rather surprisingly, everywhere in the phase diagram, indicating a kind of hidden criticality. This feature has no analog in equilibrium, and is not fully understood at present. 
 Finally, we have considered long-range open Ising models and computed squeezing numerically using matrix product states (for $N=50$ spins). While we find similar bounds on squeezing as the infinite-range model, the connection to phase transition does not appear to hold for shorter-range models. 

 It is worthwhile investigating the consequence of the quantum Fisher information saturating to $f_{\rm opt}=2$ and the possible emergence of GHZ-like states at the phase boundary. Additionally, the surprising hidden critical behavior of the mutual information well within the ordered phase is admittedly not fully understood and requires further investigation. At a technical level, developing an analytical understanding of entanglement and information measures within the ordered phase defines an interesting future direction, which will shed light on such hidden criticality.
 Finally, and most importantly for practical applications, it would be desired to identify scenarios where squeezing in the steady state can surpass the standard 3 dB limit in the presence of generic local dissipative processes such as spontaneous emission considered in this work. If possible, one could use dissipation to stabilize highly entangled states, with the clear advantage that they would be further robust against environmental noise. 
 We remark that the extensive literature on highly squeezed states has mostly focused on scenarios where only collective loss operators are present \cite{Rey-2018,Rey-2019,Clerk-2022,Schleier-Smith-2023,Shahmoon-2024}, and where the entanglement is typically optimized as a function of time. In contrast, in this work we have focused on entanglement in the steady state and in the presence of local dissipation.

\begin{acknowledgments}
We thank Aash Clerk for useful discussions. This work is supported by the Air Force Office of Scientific Research (AFOSR) under the award number FA9550-20-1-0073. We also acknowledge support from the National Science Foundation under the NSF CAREER Award (DMR2142866), as well as the NSF grants DMR1912799 and PHY2112893.  A.S.N is supported by the Dutch Research Council (NWO/OCW), as part of the Quantum Software Consortium programme (project number 024.003.037) and Quantum Delta NL (project number NGF.1582.22.030).
\end{acknowledgments}

\appendix
\section{Split system covariance matrix}\label{split system appendix}
In this appendix we show how to obtain Eq. \eqref{split cov mat}, starting from the exact field theoretical description of Eq. \eqref{iDDIM} as derived in \cite{paz_driven-dissipative_2021}. Using an exact quantum-to-classical mapping, the non-equilibrium partition function of the steady state $Z = \lim_{t \to \infty}\tr(\exp(t\mathcal{L})(\rho_0))$ can be mapped to a path integral over a pair of real fields $m_{c}, m_q$,
\begin{equation}
    Z = \int \mathcal{D}[m_c, m_q] e^{i \mathcal{S}[m_c, m_q]}\,.
\end{equation}
The action $\mathcal{S}$ is given by
\begin{equation}\label{iDDIM action}
    \mathcal{S} =  -2JN\int_t m_{c}(t)m_{q}(t) -iN\ln\tr\Big[\mathcal{T}e^{\int_t \mathbb{T}(m_{c/q}(t))}\Big]\,,
\end{equation}
where $m_c$, the ``classical" field, captures the order parameter, and $m_q$, the ``quantum" field, is related to quantum fluctuations and noise.
This exact mapping is possible due to the permutation symmetry of the Liouvillian, and through vectorization of the Liouvillian \cite{paz_driven-dissipative_2021}. The matrix $\mathbb{T}$ is a $4\times4$ matrix that can be interpreted as a transfer matrix for a pair of spins, and in the  basis where 
\begin{equation}
    \sigma^x = \begin{pmatrix}
     1 & 0 \\ 0 & -1
    \end{pmatrix}\,, \quad \sigma^y = \begin{pmatrix}
    0 & i \\ -i & 0 
    \end{pmatrix}\,,\quad \sigma^z = \begin{pmatrix}
    0 & 1 \\ 1 & 0
    \end{pmatrix}\,,
\end{equation}
it takes the form
\[
    \mathbb{T} = 
    \renewcommand*{\arraystretch}{1.5}
    \scalemath{.69}{
    \begin{pmatrix}
    -\frac{\Gamma}{4}+i2\sqrt{2}J m_q & i\Delta & -i\Delta & \frac{\Gamma}{4} \\
    i\Delta - \frac{\Gamma}{2} & -\frac{3\Gamma}{4} + i 2\sqrt{2} J m_c  & -\frac{\Gamma}{4} & -i\Delta -\frac{\Gamma}{2} \\
    -i\Delta - \frac{\Gamma}{2} &-\frac{\Gamma}{4} &- \frac{3\Gamma}{4} -i2\sqrt{2} J m_c & i\Delta - \frac{\Gamma}{2} \\
    \frac{\Gamma}{4} & -i\Delta & i\Delta & -\frac{\Gamma}{4} -i 2\sqrt{2} J m_q
    \end{pmatrix}
    }\,.
\]\par 
Connected correlation functions can be obtained by introducing source fields coupled to the desired observables to Eq. \eqref{iDDIM} \cite{paz_driven-dissipative_2021, kamenev_field_2011, sieberer_keldysh_2016}. After integration of the fields $m_c, m_q$, one obtains the generating functional $W[\{h_i\}] = i \ln Z$ which is given in terms of the desired Green's functions. We wish to obtain correlation functions for $S_x$ and $S_y$ within the same subsystem and between two different subsystems. Therefore, we define $S_{\alpha, A} = \sum_{i=1}^{N/2} \sigma_i^\alpha$ to be the collective spin operator for one half of the system, while $S_{\alpha, B} = \sum_{i=N/2}^{N} \sigma_i^\alpha$ for the other half. We introduce source fields $\mathbf{h}^{(u/l)} = (\alpha^{(u/l)}_A, \alpha^{(u/l)}_B, \beta^{(u/l)}_A, \beta^{(u/l)}_B)$ coupled to the operators $\mathbf{S}^{(u/l)} = (\sigma^{(u/l)}_{x,A}, S^{(u/l)}_{x,B}, -S^{(u/l)}_{y,A}, -S^{(u/l)}_{y,B})/\sqrt{N}$, where $u/l$ denote if the operator is acting to the left or the right of the density matrix in Eq. \eqref{iDDIM}. This modifies the vectorized Louvillian $\mathbb{L}$ (found through the transformation $A \bullet B \to A \otimes B^T = A^{(u)}B^{(l)}$ for operators $A,B$) by
\begin{equation}
    \mathbb{L}' = \mathbb{L} + i \mathbf{h}^{(u)}\cdot \mathbf{S}^{(u)} -i \mathbf{h}^{(u)}\cdot \mathbf{S}^{(u)}\,.
\end{equation}
Following the mapping to the path integral, this modifies the action in the following way,
\begin{equation}\label{split action}
\begin{split}
    \mathcal{S} =  -&2JN\int_t m_{c}(t)m_{q}(t) -i\frac{N}{2}\ln\tr\Big[\mathcal{T}e^{\int_t \mathbb{T}_A(m_{c/q}(t))}\Big] \\
    &-i\frac{N}{2}\ln\tr\Big[\mathcal{T}e^{\int_t \mathbb{T}_B(m_{c/q}(t))}\Big]\,,
\end{split}
\end{equation}
where
\begin{equation}
    \mathbb{T}_A = \mathbb{T} + i\mathbf{h}^{(u)}_A\cdot \mathbf{s}^{(u)}_A -i \mathbf{h}^{(l)}_A\cdot \mathbf{s}^{(l)}_A\,,
\end{equation}
and similarly for $\mathbb{T}_B$. The vectors with subsystem subscripts only contain fields/operators from that subsytem, and we have also defined the single spin vectors $\mathbf{s}^{(u/l)}_{A/B} = (\sigma^{x(u/l)}_{A/B}, -\sigma^{y(u/l)}_{A/B})/\sqrt{N}$.\par 
We can now expand Eq. \eqref{split action} to second order in fluctuations around $m_{c/q} = 0, \mathbf{h}_{c/q} = 0$, where we have performed the Keldysh rotation on the source fields $\mathbf{h}_{c/q} = (\mathbf{h}^{(u)} \pm \mathbf{h}^{(l)})/\sqrt{2}$. This yields an action of the form
\begin{equation}
    \mathcal{S}^{(2)} = \frac{1}{2}\int_{t, t'} \mathbf{v}^T(t) \hat{P}(t-t') \mathbf{v}(t')\,,
\end{equation}
where
\[
    \mathbf{v} = (m_c, m_q, \alpha_{c, A}, \alpha_{q,A},  \alpha_{c, B}, \alpha_{q,B},  \beta_{c, A}, \beta_{q,A}, \beta_{c, B}, \beta_{q,B})\,,
\] 
(in an abuse of notation we denote the fluctuations with the original field labels), and the kernel $P$ is a block matrix:
\begin{equation}\label{split kernel}
    \hat{P} = 
    \begin{pmatrix}
    \hat{P}_m & \hat{P}_{m , \alpha } & \hat{P}_{m , \alpha} & \hat{P}_{m, \beta} & \hat{P}_{m,\beta } \\
    \hat{P}_{\alpha, m}  & \hat{P}_\alpha & 0  & \frac{1}{2J}\hat{P}_{m, \beta} & 0\\
    \hat{P}_{\alpha, m}  & 0 &  \hat{P}_\alpha  & 0 & \frac{1}{2J}\hat{P}_{m, \beta} \\
    \hat{P}_{\beta, m} & \frac{1}{2J}\hat{P}_{\beta, m} & 0 & \hat{P}_\alpha & 0 \\
    \hat{P}_{\beta, m} & 0 & \frac{1}{2J}\hat{P}_{\beta, m} & 0 & \Hat{P}_\alpha 
    \end{pmatrix}\,.
\end{equation}
Each of the submatrices has the typical Keldysh structure,
\begin{equation}
    \hat{P}_m = \begin{pmatrix}
    0 &  P^A_m \\ P^R_m & P^K_m
    \end{pmatrix}\,.
\end{equation}
The list of elements in the time domain are
\begin{align}\label{split elements}
     P^{A/R}_m (t) &=  -2J\delta(t) + \Theta(\mp t) \left(8J^2 e^{-\frac{\Gamma}{2} | t | } \sin{(2 \Delta | t | )} \right)\,, \\
     P^K_m(t) &= i8J^2 e^{-\frac{\Gamma}{2}| t|}\cos{(2\Delta | t|)}\,, \\
     P^{A/R}_{\alpha} (t) &= \frac{1}{8J^2} \left( P^{A/R}_{m} ( t) + 2J \delta ( t) \right)\,, \\
     P^K_{\alpha} ( t) & = \frac{1}{8 J^2} P^K_m ( t)\,, \\
     P^{A/R}_{\beta,m} (t) &= \mp\Theta(\mp  t) \left( 2J  e^{-\frac{\Gamma}{2} | t | }\cos{(2\Delta | t | )}\right)\,,\\
     P^K_{\beta, m} (\delta t) &= \text{sgn}( t)\left(-i2J  e^{-\frac{\Gamma}{2}\lvert \delta t\rvert}\sin{(2\Delta \lvert\delta t\rvert)}\right)\,,
\end{align}
and the submatrices obey the following relations,
\begin{align}
    &[\hat{P}_m]^T (-t) = \hat{P}_m (t)\,,\,\,\, [\hat{P}_\alpha]^T (-t) = \hat{P}_\alpha (t)\,,\\
    &\hat{P}_{m,\alpha} (t) = 2J \hat{P}_{\alpha}\,,\,\,\,  \hat{P}_{m,\alpha} (t) = \hat{P}_{\alpha, m} (t)\,,\\ &[\hat{P}_{\beta,m}]^T (-t) = \hat{P}_{m, \beta}(t)\,.
\end{align}
\par
The final step is to integrate out the fields $m_{c/q}$ such that we are left with the generating functional. This integration is easily done in the frequency domain using functional Gaussian integration rules \cite{sieberer_keldysh_2016}, which leads to a generating functional of the form
\begin{equation}
\begin{split}
&W[\mathbf{h}] = \frac{1}{2}\int_\omega \big( \mathbf{h}^T_{\alpha_A}\hat{G}_{xx}(\omega) \mathbf{h}_{\alpha_A} + \mathbf{h}^T_{\beta_A}\hat{G}_{pp}(\omega) \mathbf{h}_{\beta_A} \\
&+ 2 \mathbf{h}^T_{\alpha_A} \hat{G}_{xp}(\omega) \mathbf{h}_{\beta_A} \big) + 2\int_\omega \big( \mathbf{h}^T_{\alpha_A}\hat{G}^{AB}_{xx}(\omega) \mathbf{h}_{\alpha_B} \\
&+ \mathbf{h}^T_{\beta_A}\hat{G}^{AB}_{pp}(\omega) \mathbf{h}_{\beta_B} +  \mathbf{h}^T_{\alpha_A} \hat{G}^{AB}_{xp}(\omega) \mathbf{h}_{\beta_B}\big) + \ldots
\end{split}
\end{equation}
where we have suppressed the frequency dependence of the fields for compactness, $\mathbf{h}_{\alpha_A} = (\alpha_{c,A}, \alpha_{q,A})$ (similarly for the other fields), and the $\ldots$ signify the rest of the terms which can be found by swapping $A \to B$. The Green's functions have a Keldysh structure,
\begin{equation}\label{greens function}
    \hat{G}_{ij} = \begin{pmatrix}
     0 & G^A_{ij} \\ G^R_{ij} & G^K_{ij}
    \end{pmatrix} \,,
\end{equation}
where $G^R_{ij}$ is the retarded response function, and the Keldysh Green's function $G^K_{ij}(t) = C_{ij}(t) = \langle \{ \delta O_i(t), \delta O_j(0)\} \rangle$ is the connected correlation function for operators $O_{i/j}$, i.e. the quantity of interest.
In terms of the submatrices given in Eq. \eqref{split kernel}, we have
\begin{align}
    \hat{G}_{xx} &= \hat{P}_{\alpha,m}(\omega) \hat{P}^{-1}_{m}(\omega) \hat{P}_{m, \alpha} (\omega) - \hat{P}_{\alpha}(\omega)\\
    \hat{G}_{pp} &= \hat{P}_{\beta,m}(\omega) \hat{P}^{-1}_{m}(\omega) \hat{P}_{m, \beta} (\omega) - \hat{P}_{\alpha}(\omega)\\
    \hat{G}_{xp} &= \hat{P}_{\beta,m}(\omega) \hat{P}^{-1}_{m}(\omega) \hat{P}_{m, \beta} (\omega) - 2\hat{P}_{\beta,\alpha}(\omega)\\
    \hat{G}^{AB}_{xx} &= \hat{P}_{\alpha,m}(\omega) \hat{P}^{-1}_{m}(\omega) \hat{P}_{m, \alpha} (\omega) \\
    \hat{G}^{AB}_{pp} &= \hat{P}_{\beta,m}(\omega) \hat{P}^{-1}_{m}(\omega) \hat{P}_{m, \beta} (\omega) \\
    \hat{G}^{AB}_{xp} &= \hat{P}_{\alpha,m}(\omega) \hat{P}^{-1}_{m}(\omega) \hat{P}_{m, \beta} (\omega) \,.
\end{align}
Taking only the Keldysh Green's function from each of these matrices, and integrating them over the frequency domain to obtain the correlation function at equal times, we retrieve the expressions for the split-system covariance matrix as shown in Eq. \eqref{split cov mat}. The identification between these results and the covariance matrix elements are
\begin{equation}
\begin{split}
    X =&\begin{pmatrix}
    C_{xx} & C^{AB}_{xx} \\ C^{AB}_{xx} & C_{xx}
    \end{pmatrix},\,
    K = \begin{pmatrix}
    C_{xp} & C^{AB}_{xp} \\ C^{AB}_{xp} & C_{xp}
    \end{pmatrix},\\
    &P = \begin{pmatrix}
    C_{pp} & C^{AB}_{pp} \\ C^{AB}_{pp} & C_{pp}
    \end{pmatrix}.
\end{split}
\end{equation}
We have used the fact that these correlation functions are symmetric under swap of $x$ and $p$.

\section{Ordered phase calculations}\label{ordered phase appendix}
Calculating the covariance matrix in the ordered phase, entire or split system, is mostly identical to how it was done in Appendix A. The main differences are how the matrix elements in Eq. \eqref{split elements} are calculated, that we also need to introduce a source for $S_z$ as there are now relevant fluctuations in the $z$-direction. In addition, we will rotate the total spin operator such that it points along the $z$-direction in the new reference frame. This is to ensure that the $\tilde{S}_x/\sqrt{2 S}$ and $\tilde{S}_y/\sqrt{2 S}$ ($S = |\mathbf{S}|$) in the new frame satisfy the canonical commutation relations, as discussed in Sec. \ref{cov mat ordered phase}. Here we will show how to perform the calculation without splitting the system in half.\par 
We begin by inserting the source field $\mathbf{h}^{(u/l)} = (\alpha^{(u/l)}, \beta^{(u/l)}, \gamma^{(u/l)})$ which couples to the spin vector $\mathbf{S}^{(u/l)} = (S^{(u/l)}_{x}, S^{(u/l)}_{y}, S^{(u/l)}_{z})/\sqrt{N}$. This modifies the matrix $\mathbb{T}$ in Eq. \eqref{iDDIM action},
\begin{equation}
    \mathbb{T}' = \mathbb{T} + i\mathbf{h}^{(u)}\cdot \boldsymbol{\sigma}^{(u)} -i \mathbf{h}^{(l)} \cdot \boldsymbol{\sigma}^{(l)}\,,
\end{equation}
where the individual spin vector is $\boldsymbol{\sigma}^{(u/l)} = (\sigma^{x(u/l)}, \sigma^{y(u/l)}, \sigma^{z(u/l)})/\sqrt{N}$.
The expansion of the exact action is now performed around the saddle-point solution, $m_c = m = \sqrt{32J\Delta - 16\Delta^2 - \Gamma^2}/4J, m_q = 0, \mathbf{h}_{c/q} = 0$, where we have performed the Keldysh rotation on the source fields $\mathbf{h}_{c/q} = (\mathbf{h}^{(u)} \pm \mathbf{h}^{(l)})/\sqrt{2}$. Defining the matrices $T_{\mu_i} = \partial_{\mu_i}\mathbb{T}$, with $\mu \in \{m, \alpha, \beta, \gamma\}, i \in \{c, q\}$, we have that
\begin{equation}
\begin{split}
    \frac{\delta^2 \log \tr (\mathcal{T}e^{\int_t \mathbb{T}})}{\delta \mu_i (t) \delta \nu_j (t')} \lvert_{\text{s.p.}} = &\Theta(t-t') \sbra{I} T_{\mu_i} e^{t \mathbb{T}_m} T_{\nu_j} \sket{\rho_{\text{ss}}}\\
    &+ \Theta(t'-t) \sbra{I} T_{\nu_j} e^{t \mathbb{T}_m} T_{\mu_i} \sket{\rho_{\text{ss}}} \\
    &- \sbra{I}T_{\mu_i}\sket{\rho_{\text{ss}}}\sbra{I}T_{\nu_i}\sket{\rho_{\text{ss}}}\,,
\end{split}
\end{equation}
where we have evaluated the left-hand side at the saddlepoint solution. We have used the fact that $\mathbb{T}$ is a vectorized single-spin Liouvillian and has a steadystate given by the vectors $\sbra{I}$ and $\sket{\rho{\text{ss}}}$. These are the left and right eigenvectors of $\mathbb{T}_m = \mathbb{T}|_{\text{s.p}}$ respectively and satisfy $\mathbb{T}_m \sket{\rho_{\text{ss}}} = 0$ (and similarly for $\sbra{I}$). We have also used the fact that $\tr(e^{\infty \mathbb{T}_m}) = 1$. The retarded elements of the expansion correspond to when $i = q, j = c$ and the Keldysh elements correspond to when $i = q, j = q$. Therefore we have generally
\begin{equation}
\begin{split}
    P^R_{\mu \nu}(t) = &-2J\delta(t) \delta_{\mu, m}\delta_{\nu, m} \\
    &- iN\Theta(t)\sbra{I} T_{\mu_q} e^{t \mathbb{T}_m} T_{\nu_c} \sket{\rho_{\text{ss}}}
\end{split}
\end{equation}
\begin{equation}
\begin{split}
    P^K_{\mu \nu}(t) = &-iN\Theta(t) \sbra{I} T_{\mu_q} e^{t \mathbb{T}_m} T_{\nu_q} \sket{\rho_{\text{ss}}}\\
    &-iN\Theta(-t) \sbra{I} T_{\nu_q} e^{t \mathbb{T}_m} T_{\mu_q} \sket{\rho_{\text{ss}}}\,.
\end{split}
\end{equation}
The fact that $m$ is finite makes further expansion of the trace-log in Eq. \eqref{iDDIM action} difficult analytically \cite{paz_driven-dissipative_2021}. 
To simplify this expression, we can rewrite the generator as
\begin{equation}
    e^{t \mathbb{T}_m} = \sket{\rho_{\text{ss}}}\sbra{I} + \sum_{k =1}^3 e^{t \lambda_k} \sket{\lambda_k^R}\sbra{\lambda_k^L}\,,
\end{equation}
where each of the eigenvalues $\lambda_k$ are either real or come in complex conjugate pairs with $\text{Re}(\lambda_k)<0$. We then Fourier transform the result, using the definition $\phi(t) = \frac{1}{2\pi}\int d\omega e^{-i\omega t}\phi(\omega)$ for any field $\phi(t)$. We find
\begin{equation}
    P^R_{\mu\nu}(\omega) = -2J\delta_{\mu, m}\delta_{\nu, m} +i N \sum_{k=1}^3 D_{\mu \nu}(k) \frac{1}{\lambda_k + i \omega} 
\end{equation}
\begin{equation}
    P^K_{\mu \nu}(\omega) = iN\sum_{k = 1}^3 \tilde{D}_{\mu \nu}(k) \frac{1}{\lambda_k + i \omega} +  \tilde{D}_{\nu \mu}(k) \frac{1}{\lambda_k - i \omega}\,,
\end{equation}
where we have defined the coefficients
\begin{align}
    D_{\mu\nu}(k) &= \sbra{I}T_{\mu_q}\sket{\lambda^R_k}\sbra{\lambda^L_k}T_{\nu_c}\sket{\rho_{\text{ss}}}\\
    \tilde{D}_{\mu\nu}(k) &= \sbra{I}T_{\mu_q}\sket{\lambda^R_k}\sbra{\lambda^L_k}T_{\nu_q}\sket{\rho_{\text{ss}}}\,,
\end{align}
for compactness.
These are the matrix elements of the submatrices in a block matrix similar to that of Eq. \eqref{split kernel}, except now there are no subsystems and we have new blocks from the $\gamma$ source field. Following that procedure, we arrive at the Green's functions in the ordered phase, 
\begin{equation}
\begin{split}
    \hat{G}_{\mu \nu} = &\hat{P}_{\mu \nu} (\omega) \hat{P}^{-1}_{mm}(\omega) [\hat{P}_{\nu \mu}]^T(-\omega)\\
    &- 2\hat{P}_{\mu \nu}(\omega) + \delta_{\mu,\nu}\hat{P}_{\mu \mu}\,,
\end{split}
\end{equation}
where $\mu,\nu \in \{\alpha, \beta, \gamma\}$ as we have eliminated the $m$ fields. The correlation functions for $S_{x,y,z}/\sqrt{N}$ are then all given by 
\begin{equation}
    C_{\mu \nu} = \int_\omega G^K_{\mu \nu}(\omega)\,,
\end{equation}
where the Keldysh component is as defined in Eq. \eqref{greens function}. The labels $\mu,\nu$ can now be identified with $x,y,z$ instead of $\alpha, \beta, \gamma$ respectively. As mentioned before, it is difficult to obtain an analytical expression for the correlation functions. However, one can numerically evaluate the coefficients $D_{\mu \nu},\tilde{D}_{\mu\nu}$ and then integrate $G^K_{\mu\nu}$ over its $\omega$ dependence to obtain a numerical value for the correlation functions.\par 
To obtain a representation in terms of canonical variables as we did in the normal phase, we rotate the spin observables such that $\tilde{S}_z$, the rotated spin operator, points along $\mathbf{n} = \mathbf{S}/|\mathbf{S}|$. This is done using the transformation matrix \cite{lerose_impact_2019}
\begin{equation}
    R(\theta, \phi) = \begin{pmatrix}
    \cos{\theta}\cos{\phi} & \cos{\theta}\sin{\phi} & -\sin{\theta} \\
    -\sin{\phi} & \cos{\phi} & 0 \\
    \sin{\theta} \cos{\phi} & \sin{\theta}\sin{\phi} & \cos{\theta}
    \end{pmatrix}\,,
\end{equation}
which defines the new spin variables
\begin{equation}
    \tilde{\mathbf{S}} = R(\theta, \phi) \mathbf{S}\,.
\end{equation}
The angles $\theta, \phi$ that achieve the desired rotation are given by
\begin{equation}
    \theta = \cos^{-1}\left( \frac{Z}{s}\right)\,,\,\,\,   \phi = \cot^{-1}\left(\frac{X}{Y}\right)\,,
\end{equation}
where $X,Y,Z$ are the mean-field solutions defined in Eq. \eqref{mean-field eqs}, and $s = \sqrt{X^2 + Y^2 + Z^2}$. This rotation gives the new spin operators in terms of the old ones, which means we also know what the correlation functions of the new variables are in terms of the old ones.

\end{document}